\documentclass[useAMS,usenatbib]{mnras}
\usepackage{graphicx}
\usepackage{epstopdf}
\usepackage{enumerate}
\usepackage{amsmath}
\usepackage{array}
\usepackage[T1]{fontenc}
\def\be{\begin{equation}}
\def\ee{\end{equation}}
\usepackage{float}
\usepackage[usenames,dvipsnames,svgnames,Table]{xcolor}

\title[Broadband spectral modelling of GX~339$-$4]{Combining timing characteristics with physical broadband spectral modelling of black hole X-ray binary GX~339$-$4}

\author[R. M. T. Connors et al.]
    {R.~M.~T.~Connors$^{1,\,4}$\thanks{E-mail: rconnors@caltech.edu},
       D.~van Eijnatten$^1$, S.~Markoff$^{1,\,2}$, C.~Ceccobello$^{1,\,6}$, 
      \newauthor V.~Grinberg$^3$, L.~Heil$^1$\thanks{No longer affiliated - Email: lucyheil@gmail.com }, D.~Kantzas$^{1,\,2}$, M.~Lucchini$^1$, P.~Crumley$^{1,\,5}$\\
     $^1$Anton Pannekoek Institute, University of Amsterdam, Science Park 904, 1098 XH Amsterdam, The Netherlands\\
     $^2$GRAPPA, University of Amsterdam, Science Park 904, 1098 XH Amsterdam, The Netherlands\\
     $^3$Institut f\"ur Astronomie und Astrophysik (IAAT), Universit\"at T\"ubingen, Sand 1, 72076 T\"ubingen, Germany\\
     $^4$Cahill Center for Astronomy and Astrophysics, California Institute of Technology, 1200 California Boulevard, Pasadena, CA 91125, USA\\
     $^5$Department of Astrophysical Sciences, 4 Ivy Lane, Princeton University, Princeton, NJ 08544 \\
     $^6$Chalmers University of Technology, SE-412 96, Gothenburg, Sweden}
     \date{}

\pagerange{\pageref{firstpage}--\pageref{lastpage}} \pubyear{2019}

\begin{document}

\label{firstpage}

\maketitle

\begin{abstract}
\\
GX~339$-$4 is a black hole X-ray binary that is a key focus of accretion studies since it goes into outburst roughly every two-to-three years. Tracking of its radio, IR and X-ray flux during multiple outbursts reveals tight broadband correlations. The radio emission originates in a compact, self-absorbed jet, however the origin of the X-ray emission is still debated: jet base or corona? We fit 20 quasi-simultaneous radio, IR, optical and X-ray observations of GX~339$-$4 covering three separate outbursts in 2005, 2007, 2010--2011, with a composite corona + jet model, where inverse Compton emission from both regions contributes to the X-ray emission. Using a recently-proposed identifier of the X-ray variability properties known as power-spectral hue, we attempt to explain both the spectral and evolving timing characteristics, with the model. We find the X-ray spectra are best fit by inverse Compton scattering in a dominant hot corona ($kT_{\rm e}\sim$ hundreds of keV). However, radio and IR-optical constraints imply a non-negligible contribution from inverse Compton scattering off hotter electrons ($kT_{\rm e} \ge 511$~keV) in the base of the jets, ranging from a few up to $\sim50$\% of the integrated 3--100~keV flux. We also find that the physical properties of the jet show interesting correlations with the shape of the broadband X-ray variability of the source, posing intriguing suggestions for the connection between the jet and corona.
\end{abstract}


\section{Introduction}
\label{sec:intro}

Accreting black holes are found to exist across a wide range of masses, from the stellar-mass remnants of stars (discoverable as the primaries of binary systems; black hole X-ray binaries, from here on BHBs), to their supermassive ($10^6$--$10^{10}~M_{\odot}$) analogues at the centres of galaxies (active galactic nuclei; AGN). Despite the disparity in mass, size-scale, and local environment, there is growing evidence that the physical nature of accretion flows around stellar-mass and supermassive black holes is mass-invariant, at least in the innermost regions (e.g., \citealt{Merloni2003,F04,kjf06,Plotkin2012}). This apparent scale-invariant property of accretion has led to the hypothesis that the diversity of AGN types arises due to the combination of observer viewing angle (e.g., \citealt{Urry1995}) and the evolving states of AGN (e.g., \citealt{Merloni2003,F04,kjf06}), akin to the changes we see occurring in BHBs (see, e.g., \citealt{Nowak1995,vanderKlis1995,RemMc2006,bel10}). However, tracking the long timescale evolution of individual AGN to observe such state changes is not possible due to the orders-of-magnitude difference in dynamical times compared to those of BHBs. We can instead further our understanding of the evolving properties of BHBs, determine the physical conditions under which state-changes occur, and then see if it can account for the phenomenology observed in different AGN types. 
 
 The spectral classification of BHB states can be loosely divided into two categories: soft and hard (see, e.g., \citealt{Nowak1995} for a review). In soft BHB states the X-ray spectrum is dominated by a soft (peak at $\sim1~\mathrm{keV}$) multi-temperature blackbody component, attributable to optically thick emission originating from a thin accretion disc \citep{SS1973}. In hard BHB states we instead see a spectrum dominated by hard power-law emission which has a more ambiguous origin. Some models adopt either static or inflow geometries, whilst others place the emission region within an outflow/jet. Inflow/static models include inverse Compton (IC) or synchrotron self-Compton (SSC) scattering within an optically thin `corona' or radiatively inefficient accretion flow (RIAF) in the inner regions of the accretion flow \citep{Lightman1974,Eardley1975,Shapiro1976,Haardt1993, ny94,Narayan1995,Esin1997}. Outflow models instead propose either SSC/IC or optically thin synchrotron from within a jet/outflow \citep{Markoff2001,mnw05, Yuan2005,Romero2008}. Understanding the interplay between these spectral components, determining which is the dominant mechanism at play, and explaining the connection between the accretion disc, corona and jet, is a key focus of recent targeted multiwavelength observing campaigns on BHBs (e.g. \citealt{Corbel2000,Corbel2003,Gandhi2008,Gandhi2010,Miller-Jones2012,Russell2013,Corbel2013,Russell2014}). X-ray variability studies show state classifications can also be made in the time domain, with low-rms/high-rms variability observed in soft/hard states respectively (see, e.g., \citealt{rm06} for a review). A full outlook on the structure and evolution of BHBs comes from studies in both the spectral the time-variability domains.

 Targeted observing campaigns focused on BHB outbursts have led to an empirical correlation between their X-ray and radio fluxes (e.g. \citealt{Hannikainen1998,Corbel2000,Corbel2003,gfp03,Corbel2008,MillerJones2011,Corbel2013,Gallo2014}), and these correlations have been extended into the optical/NIR bands (e.g. \citealt{Russell2006}). The radio/X-ray correlation has been observed to cover several orders of magnitude in luminosity in the low hard states of some sources, such as GX~339$-$4 \citep{Corbel2000,Corbel2003,Corbel2013}, and V404~Cygni \citep{Corbel2008}, and the sources track the same correlation over different outbursts. So the radio/X-ray correlation is locked in as BHBs evolve through their hard states. These correlations indicate that the allocation of power between the physical components in BHBs, as a function of accretion rate, is an intrinsic property of hard state BHBs. In order to draw robust conclusions about the drivers of state changes and the nature of jet launching, we need to be able to reliably identify the source of the X-ray emission, and determine the exact nature of the connection between the X-ray-emitting regions and the jet radio core.

\begin{figure*}
\vspace{-2.5cm}
\includegraphics[width=0.6\linewidth,angle=270]{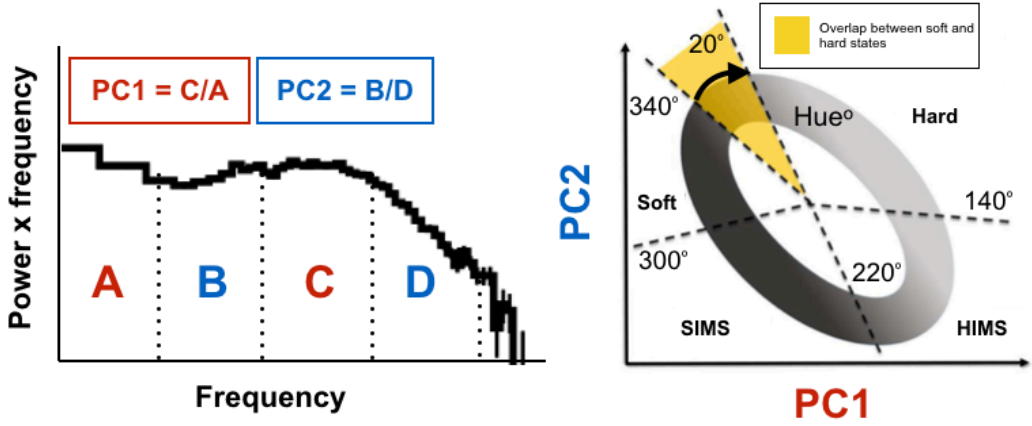}
\vspace{-2.5cm}
\caption{\textbf{Left}: A conceptual diagram representing a BHB power spectrum divided into frequency bins in log space. Two power-colour ratios are defined as PC1 = C/A, PC2 = B/D, where A, B, C, and D are the integral power across the defined frequency bands. \textbf{Right}: The power-colour hue diagram taken from Heil et al. (2015). The angular position in degrees (where $0^{\circ}$ corresponds to the semi-major axis at $45^{\circ}$ to the x-and-y axes) is defined as the hue, and the corresponding states are marked roughly; Soft, Hard, HIMS (hard intermediate state) and SIMS (soft intermediate state). Soft and hard states overlap in the top left of the diagram because their power-spectra have a similar shape, though the normalisations are different---hard states have stronger broad-band variability than soft states. A BHB will start from the top left of the diagram, follow a clockwise path during outburst back to its original position, and then move anti-clockwise through outburst decay back towards the hard state.}
\label{fig:heil}
\end{figure*}

 
 Further developing models of the accretion flow and how it interacts with the jet/outflow in BHBs requires a combination of broadband (radio-to-X-ray) timing and spectral information; these two pictures are seldom treated in unison however, unfortunately, despite the wealth of variability phenomena. However, \cite{Heil2015} developed a novel state classification method for BHBs which characterises the shape of the power spectrum of their X-ray light curves through the course of an outburst, analogous with the well-known hardness intensity diagram of BHB states (HID; \citealt{Homan2001,Homan2005,Belloni2004,Belloni2005}). A single variable, the power-spectral `hue', encodes the relation between two ratios of integrated power across individual frequency bands in Fourier space (see Figure~\ref{fig:heil}). One can use this information to track spectral properties alongside timing characteristics. Since the timing characteristics represent complementary changes in the system configuration over time, we would expect to see some consistency between the physical state of the inner accretion flow/jet and the hue. 

In this paper we combine, for the first time, the X-ray variability classification scheme of \cite{Heil2015} with broadband spectral information to build a consistent picture of the evolution of the jet and inner accretion flow of GX~339$-$4. By probing the dominant spectral components and comparing model parameters with the evolution of its variability, we develop a somewhat quantitative description of changes to the accretion flow and jet during both the rise and decay of its outburst. We focus in particular on the relative dominance of the jet and corona in the X-ray band. In Section \ref{sec:gx339anddata} we present the radio-IR-optical-X-ray data compilation we use for model-fitting. In Section \ref{sec:model} we briefly discuss the outflow-dominated model used in our fits. In Section \ref{sec:method} we present our spectral-modelling method and the results of fits to X-ray and broadband (radio, IR-optical, X-ray) spectra, as well as the key parameter trends with variability properties of GX~339$-$4, and a brief consideration of high-energy pair processes in the jet. In Section \ref{sec:discussion} we discuss the significance of these parameter trends and comparisons with previous modelling of the broadband spectra GX~339$-$4. In Section \ref{sec:conclusions} we summarise our results and conclude.

\section{GX~339$-$4: physical characteristics and data selection}
\label{sec:gx339anddata}
GX~339$-$4 has been one of the most intensely studied BHBs since its discovery in 1973 \citep{Markert1973}, due primarily to its short X-ray duty cycle (going into outburst roughly every 2--3 years). As such we have extensive spectral and timing information of GX~339$-$4 covering multiple outbursts (7 with simultaneous radio/X-ray coverage; see \citealt{Corbel2013}), making it the ideal candidate for studies of how spectral properties (and the physical mechanisms behind them) track the time variability behaviour in BHBs. \\
\indent One caveat of conducting such studies on GX~339$-$4 is the lack of accuracy achieved in determining its physical properties. The most heavily cited and utilised mass function measurement is that obtained by \cite{Hynes2003} of $5.8\pm0.8~M_{\odot}$, and a later estimate included a lower limit of $7~M_{\odot}$ \citep{Munoz-Darias2008}. In contrast more recent near-infrared detections of absorption lines from the donor star of GX~339$-$4 indicate a mass function of $\sim1.91\pm0.08~M_{\odot}$ \citep{Heida2017}. Distance has also been difficult to determine\footnote{Though the recent Gaia survey \citep{Gaia} has already led to new distance estimates of many BHBs \citep{Gandhi2018}}, with early estimates finding a broad range from 6--15 kpc \citep{Hynes2004}, and best estimates giving $\sim8~\mathrm{kpc}$ \citep{Zdziarski2004}, based on a comparison of the redshifted spectral lines seen in GX~339$-$4 with those of stars in the Galactic bulge region at $D=8\pm2~\mathrm{kpc}$, and the high peculiar velocity of GX~339$-$4 ($v\sim 140~\mathrm{km~s^{-1}}$; \citealt{Hynes2004}). One of the most elusive physical properties of GX~339$-$4 has been the orbital inclination. With almost no model-independent consensus, we mostly rely on modelling of accretion disc reflection in the X-ray spectra to determine the inclination. Reflection modelling has derived inclination estimates over a large range: 15$^{\circ}$--50$^{\circ}$ \citep{Miller2006,Reis2008,Done2010,Plant2014,Plant2015,Garcia2015,Parker2016}. These values are not wholly reliable for two reasons: 1) these are model-dependent estimates that are degenerate with other key parameters of the reflection models, and 2) the disc inclination may not be equal to the orbital inclination of the binary (see, e.g., \citealt{Wijers1999,Maccarone2002,Begelman2006}). We nonetheless adopt the best estimates possible in order to model the data. We choose to fix the observational characteristics of GX~339$-$4 at distance $D=8~\mathrm{kpc}$ \citep{Zdziarski2004}, inclination $i=40^{\circ}$ (a rough average of the broad range of estimates), and mass $M_{\mathrm{BH}}=f(M)(1+q)^2/\sin^3i=9.8~M_{\odot}$, adopting the mass function of \cite{Heida2017} and assuming the mass of the donor star to black hole mass is $q=0.17$. \\
\indent GX~339$-$4 has a compact radio jet during the hard state \citep{Fender2001}, and the emission from this jet dominates up to IR \citep{Corbel2002} and possibly optical frequencies \citep{Gandhi2008,Gandhi2010,Gandhi2011,Casella2010}. Correlations between the optical/IR/X-ray light curves during various GX~339$-$4 outbursts indicate a physical connection between the regions near the black hole and the self-absorbed regions of jets at $\sim10^3$--$10^4~r_{\rm g}$, supported by recent optical and IR lags of $\sim100$~ms (with respect to X-ray) detected from the jet \citep{Gandhi2010,Kalamkar2016}; a roughly equivalent lag was detected in BHB V404~Cygni recently too \citep{Gandhi2017}. This lag between emission at high and low frequencies in BHBs is best interpreted as variations propagating through the jet. Such variations are thought to be associated with accretion rate fluctuations propagating through the disc \citep{Lyubarskii1997,Uttley2001}.


\subsection{Data}
 \label{sec:data}
 We compile data from 20 separate quasi-simultaneous (all observations within 24-hrs of one another), broadband observations of GX~339$-$4, covering the radio, near-IR/optical, and X-ray bands. Here we describe how the data were collected and reduced. In basic terms, the selection criteria are that there is quasi-simultaneous broadband coverage of GX~339$-$4 and that it is in its hard state, defined by its variability and spectral properties (hue and hardness ratio). 
 
\subsubsection{X-ray data}
\begin{figure}
 \includegraphics[width=\linewidth]{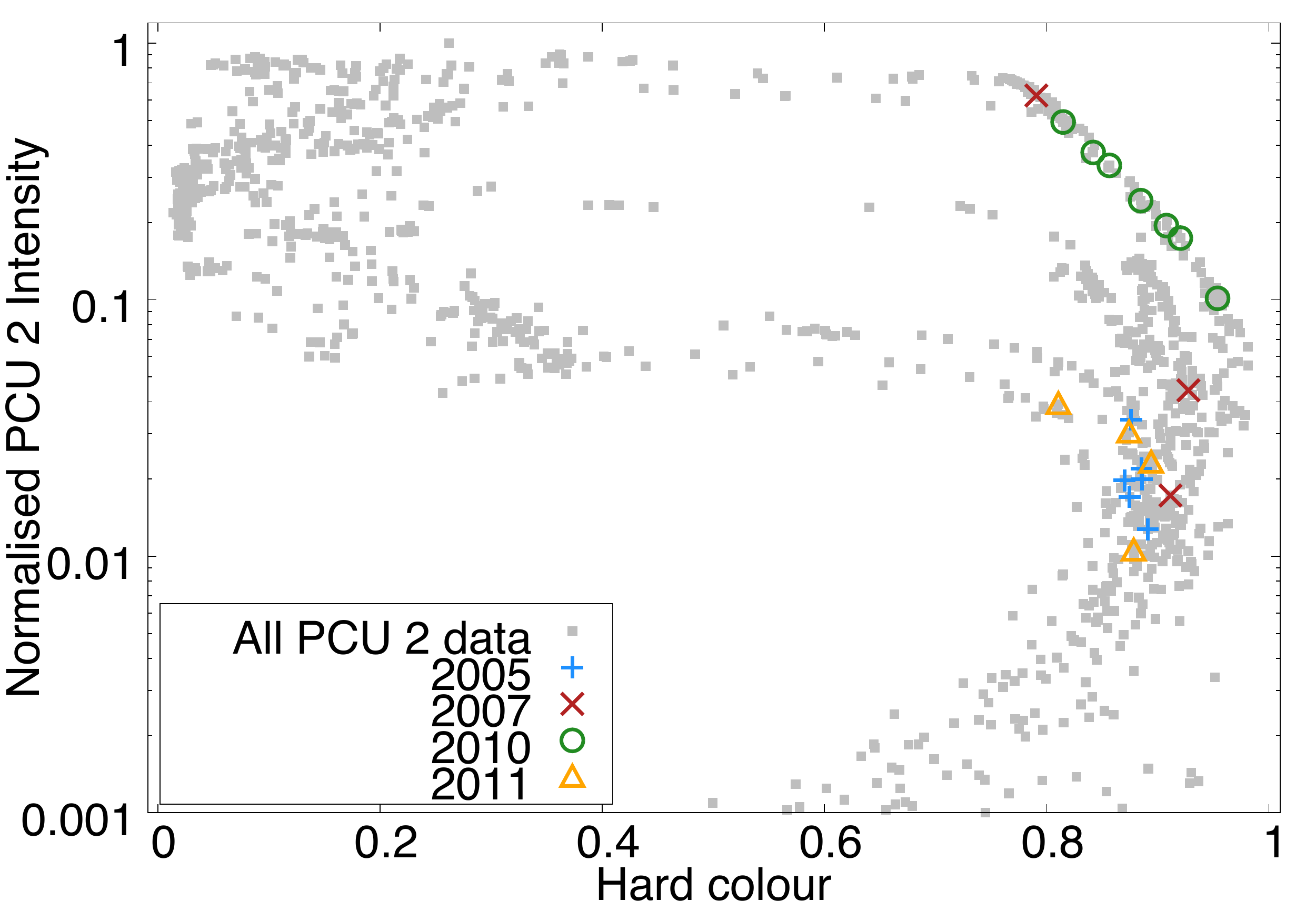}\vspace{-0.5cm}
 \caption{Hardness-Intensity diagram showing all the \textit{RXTE}-PCU 2 observations of GX~339$-$4 (gray points) and highlighting the 20 observations in our sample, divided by observation year, with the colours/symbols indicated in the key. Hardness ratios and intensities are calculated using all Xenon-layers of the PCU~2 detector (as opposed to utilizing only the top Xenon layer as in the sample of modelled data) in order to show absolute values in line with ratio measurements in the literature. Intensities are normalised by the peak average intensity of the source. Hard colour is defined as the ratio of source counts in the 8.6--18~keV to 5--8.6~keV energy bands respectively.}
 \label{fig:HID}
 \end{figure}

Data from the \textit{Rossi X-ray Timing Explorer (RXTE)} proportional counter array \citep[PCA]{Jahoda_2006a} and High-Energy Timing Experiment \citep[HEXTE]{Rothschild_1998a} were extracted using HEASOFT 6.16 following the standard procedure as described, e.g., in \cite{Grinberg_2013a}, in particular discarding data within 10 minutes of the South Atlantic Anomaly passages.\\
\indent For the PCA, we use data from the top xenon layer of proportional counter unit (PCU)~2 only since these data are best calibrated. We apply \texttt{PCACORR} calibration tool \citep{Garcia_2014a} to further improve the data quality. No HEXTE data are available for over half of our observations (Table~\ref{tab:data}) due to the failure of the rocking mechanisms of both HEXTE clusters late in the \textit{RXTE} mission lifetime. We extract cluster A and B data where available. We refrain from using the \texttt{HEXTECORR} calibration tool \citep{Garcia_2016a} on the HEXTE~B data as the improvement would only be marginal given the data quality.\\
\indent The PCA light curves are used to calculate the power-spectral hue of each observation (shown in Table~\ref{tab:data}), following the method of \cite{Heil2015}. Figure~\ref{fig:heil} shows how the PCU~2 light curves are used to calculate the power-spectral hue. A fourier transform is taken, and the resulting power spectrum is divided into four roughly even log-spaced frequency bands: A = 0.0039--0.031~Hz, B = 0.031--0.25~Hz, C = 0.25--2.0~Hz, D = 2--16~Hz. The ratios of integrated power between bands C/A and B/D are then taken, defining power-colour 1 (PC1) and power-colour 2 (PC2) respectively. Placed on a scatter plot of PC1 and PC2, the data follow an annulus. The clockwise angular position of each observation (with respect to a semi-major axis at 45$^{\circ}$ to the x-and-y axes) defines its power-colour hue. All our X-ray data have been pre-selected with $-20^{\circ}<$~hue~$<140^{\circ}$, which \cite{Heil2015} define as the hard state, and all have a hardness ratio $>0.75$. The corresponding hardness ratios for all the PCU~2 spectra are shown in Table~\ref{tab:data} and displayed in a hardness-intensity diagram in Figure~\ref{fig:HID} for clarity, although we note that those values have been calculated using counts collected by all 3 Xenon layers of the detector. 

  \begin{figure}
  \vspace{-0.5cm}
 \includegraphics[width=\linewidth]{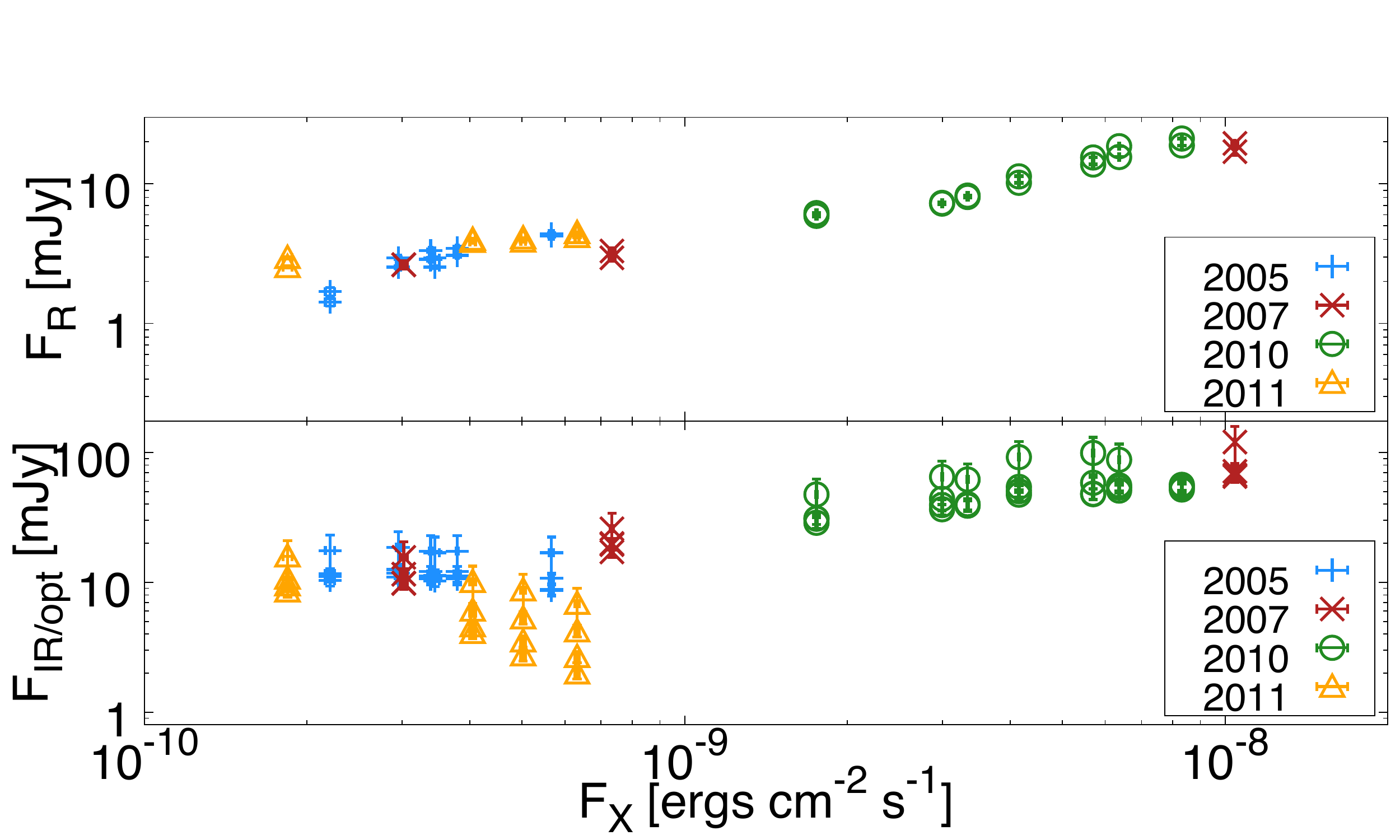}
 \vspace{-0.5cm}
 \caption{The radio (5.5~GHz and 8.8~GHz) and OIR (V, I, J and H bands) fluxes of all 20 quasi-simultaneous broadband observations of GX~339$-$4 in mJy alongside the X-ray PCA data fluxes [3--20~keV] in erg~cm$^{-2}$~s$^{-1}$. The fluxes are separated by observation years 2005, 2007, 2010 and 2011, covering three separate outbursts, with the key indicating the symbols and colours corresponding to each year.}
 \label{fig:data_fluxes}
 \end{figure}
 
 \begin{figure}
 \centering
 \hspace{-0.5cm}
 \includegraphics[width=1.05\linewidth]{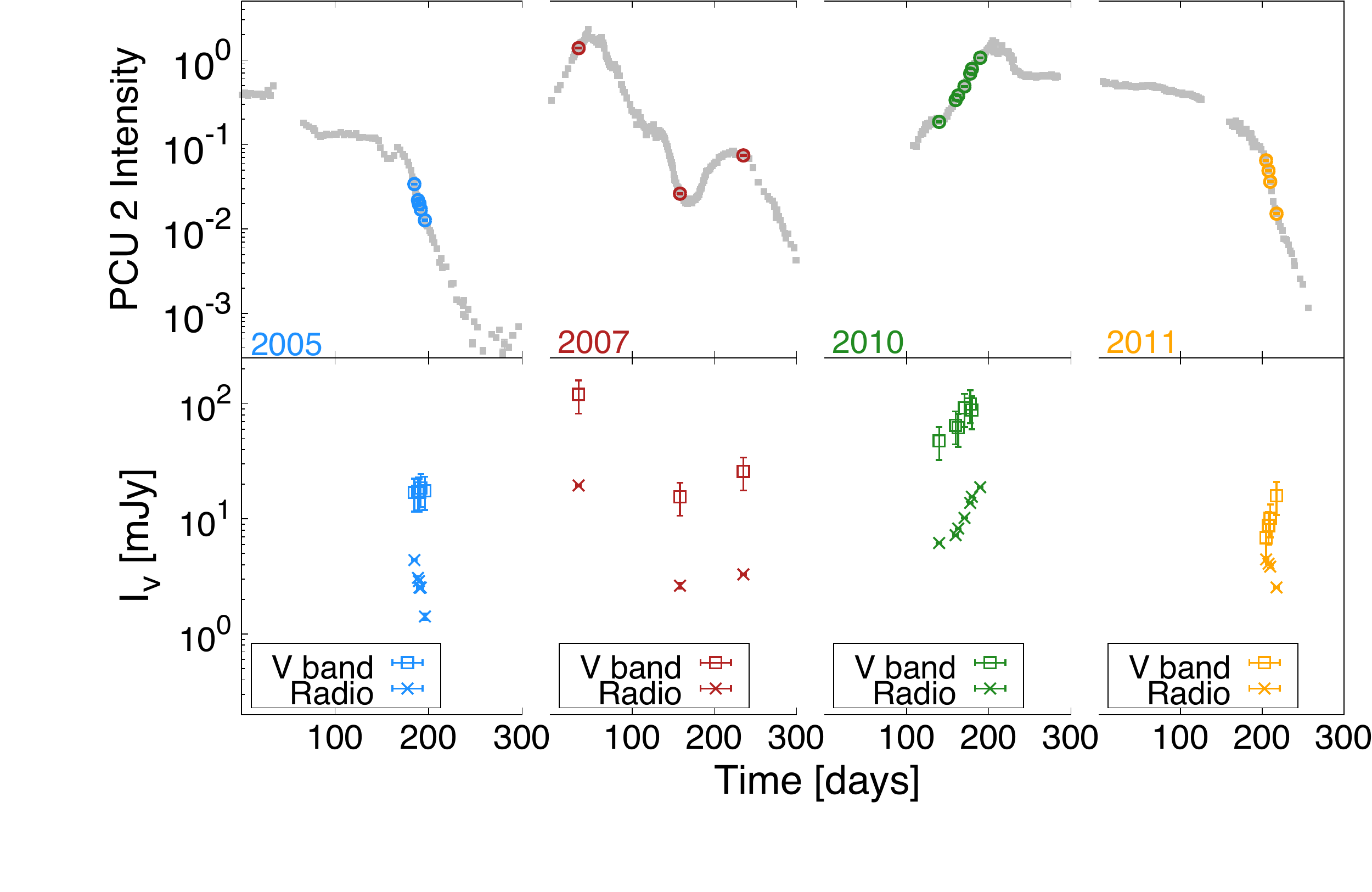}\vspace{-0.8cm}
 \caption{The PCU~2 intensity (top), optical V band, and radio fluxes (bottom) as a function of MJD, divided by observation year, truncated in time to fit together on one plot. Each panel shows a 300 day snapshot of the PCU~2 light curve, with the following MJD ranges for each year: 53300--53600 (2005), 54100--54400 (2007), 55100--55400 (2010), 55400--55700 (2011).}
 \label{fig:lcurves}
 \end{figure}

\subsubsection{Radio/IR-Optical data}
\label{subsubsec:oir}
We select radio fluxes of GX~339$-$4 covering a 15-year period (1997--2012) resulting from observations made with the Australian Telescope Compact Array (ATCA) \citep{Corbel2013}, choosing only those observations falling within a 24-hr window of the corresponding X-ray observations. We then include optical and near-infrared fluxes resulting from observations of GX~339$-$4 made with the SMARTS 1.3 m telescope from 2002--2010, covering the \textit{V}, \textit{J}, \textit{I} and \textit{H} bands \citep{Buxton2012}. The magnitudes in all four bands are de-reddened assuming $n_\mathrm{H}=5\pm1\times10^{21}~\mathrm{cm}^{-2}$ \citep{Kong2002}, giving $E(B-V)=0.94\pm0.19$ \citep{PS1995}, such that $A_V=2.9\pm0.6$ \citep{Cardelli1989}. The flux density values quoted in Table~\ref{tab:data} are the de-reddened flux densities given by \cite{Buxton2012}. We reject SMARTS observations that fall outside the 24 hour window of the pre-selected quasi-simultaneous radio and X-ray observations. This selection criterion leaves us with 20 separate broadband quasi-simultaneous spectra of GX~339$-$4, covering the decay of its 2005 outburst, the peak and decay of its 2007 outburst, and the rise and decay of its 2010 outburst. A full description of the data is shown in Table~\ref{tab:data}, and see Figure~\ref{fig:data_fluxes} for a plot of the radio and IR/optical (OIR) fluxes against the X-ray fluxes of all 20 observations, and Figure~\ref{fig:lcurves} for X-ray lightcurves with optical and radio band fluxes showing the different outburst stages our datasets probe. One notices instantly that the first three observations during the 2011 outburst decay have notably lower OIR fluxes for their given X-ray fluxes than in all other observations, with a trend that deviates from an otherwise well-behaved correlation. \\
\indent Selecting quasi-simultaneous data with a 24-hr time-window coincidence across radio/OIR/X-ray bands in this way optimises the trade-off between the quantity of data we require for our modelling, and the information lost by neglecting source variability on short timescales. \cite{Gandhi2011} show that the mid-IR spectral slope is variable on timescales of $\sim20~\mathrm{minutes}$. We therefore highlight the uncertainties in the overall flux and spectral slope incurred by grouping data over the 24-hr time window, and simply note it as a caveat to our analysis.

\begin{table*}
\centering
\caption{The broadband quasi-simultaneous data of GX~339$-$4. Shown from left to right: (1) spectrum number, (2) MJD, (3) the 5.5~GHz and 8.8~GHz radio fluxes, (4) the OIR fluxes, bands $V$, $I$, $J$, and $H$, (5) the observational ID of the \textit{RXTE} observation, (6) the X-ray unfolded PCA data flux (model indepedent), 3--20~keV, (7) the hardness ratio, defined as the ratio of PCU~2 (all layers) source counts between the [8.6--18~keV]/[5--8.6~keV] bands, (8) the power-spectral hue, (9) HEXTE cluster A or B spectra included.}
\label{tab:data}
\begin{tabular}{@{}lcccccccc}
\hline
Spec. \# & MJD & $F_{\rm R}$ [mJy] & $F_{\rm IR/opt}$ [mJy] & ObsID & $F_{\rm X}$  & HR & Hue & HEXTE?\\
& (-245000) & 5.5 GHz & $V, I$  & & [$10^{-10}~\mathrm{erg}~\mathrm{s}^{-1}~\mathrm{cm}^{-2}$] & & [$^{\circ}$]\\
& & 8.8 GHz & $J, H$ & & [3--20 keV]\\
\hline
1 & 53485 & $4.39\pm0.06$  & $17\pm5$, $11\pm1$  & 90704-01-13-01 & $5.67\pm0.05$ & 0.88 & $42\pm13$ & A \& B\\
    &           & $4.23\pm0.08$ & $8.9\pm0.9$, $8.7\pm0.8$\\
2 & 53489 & $3.1\pm0.1$  & $17\pm6$, $12\pm1$  &  91095-08-06-00 & $3.79\pm0.02$ & 0.89 & $32\pm4$ & A \& B \\
   &            & $3.5\pm0.1$ & $11\pm1$, $11\pm1$\\
3 & 53490 & $2.88\pm0.08$  & $17\pm6$, $12\pm1$  &  91095-08-07-00 & $3.39\pm0.02$ & 0.87 & $22\pm3$ & A \& B\\
   &            & $3.3\pm0.1$ & $11\pm1$, $11\pm1$\\
4 & 53490 & $2.53\pm0.05$  & $17\pm5$, $11\pm1$  & 91105-04-17-00 & $3.45\pm0.07$ & 0.89 & $20\pm10$ & No\\
   &            &  $2.94\pm0.07$ & $11\pm1$, $10\pm1$\\
5 & 53492 & $2.53\pm0.05$  & $19\pm6$, $13\pm1$   & 91095-08-09-00 & $2.95\pm0.03$ & 0.87 & $24\pm4$ & A \& B\\
   &            & $2.94\pm0.07$ & $12\pm1$, $11\pm1$\\
6 & 53496 & $1.42\pm0.09$  & $18\pm6$, $12\pm1$   & 90704-01-14-00 & $2.21\pm0.04$ & 0.89 & $9\pm6$ & No\\
   &            & $1.7\pm0.1$ & $11\pm1$, $10\pm1$\\
7 & 54135 & $19.5\pm0.3$  & $120\pm40$, $72\pm7$  & 92035-01-02-02 & $104.2\pm0.1$ & 0.79 & $87\pm5$ & B only\\
   &            & $17\pm1$ & $67\pm6$, $65\pm6$\\
8 & 54258 & $2.6\pm0.2$  & $16\pm5$, $12\pm1$   & 92704-03-26-00 & $3.02\pm0.05$ & 0.91 & $22\pm10$ & No\\
   &            & $2.6\pm0.2$ & $10\pm1$, $10\pm1$\\
9 & 54335 & $3.3\pm0.05$  & $26\pm8$, $20\pm2$  & 93409-01-05-03 & $7.33\pm0.05$ & 0.93 & $18\pm6$ & B only\\
   &            & $2.95\pm0.07$ & $17\pm2$, $20\pm2$\\
10 & 55240 & $6.17\pm0.06$  & $47\pm15$, $29\pm3$  & 95409-01-06-00 & $17.53\pm0.05$ & 0.95 & $347\pm3$ & No\\
   &            & $5.9\pm0.1$ & $31\pm3$, $31\pm3$\\
11 & 55260 & $7.2\pm0.1$  & $65\pm21$, $44\pm4$  & 95409-01-08-03 & $29.94\pm0.06$ & 0.92 & $359\pm2$ & No\\
   &            & $7.3\pm0.1$ & $37\pm4$, $39\pm4$\\
12 & 55263 & $8.24\pm0.05$  & $62\pm20$, $39\pm4$  & 95409-01-09-01 & $33.4\pm0.1$ & 0.91 & $358\pm3$ & No\\
   &            & $8.1\pm0.1$ & $40\pm4$, $39\pm4$ \\
13 & 55271 & $10.2\pm0.1$  & $92\pm29$, $54\pm5$  & 95409-01-10-03 & $41.53\pm0.07$ & 0.88 & $15\pm3$ & No\\
   &            & $11.3\pm0.1$ & $50\pm5$, $47\pm5$\\
14 & 55277 & $13.8\pm0.1$  & $99\pm32$, $58\pm6$  & 95409-01-11-02 & $57.0\pm0.2$ & 0.86 & $19\pm13$ & No\\
   &            & $15.45\pm0.06$ & $48\pm5$, $48\pm5$\\
15 & 55280 & $15.56\pm0.05$  & $88\pm28$, $56\pm5$  & 95409-01-11-03 & $63.7\pm0.2$ & 0.84 & $45\pm10$ & No\\
   &            & $18.59\pm0.05$ & $54\pm5$, $51\pm5$\\
16 & 55290 & $18.8\pm0.1$  & $N/A$, $54\pm5$  & 95409-01-13-00 & $83.2\pm0.1$ & 0.81 & $38\pm26$ & No\\
   &            & $21.1\pm0.2$ & $56\pm5$, $52\pm5$\\
17 & 55605 & $4.45\pm0.04$  & $7\pm2$, $4.2\pm0.4$  & 96409-01-07-03 & $6.32\pm0.05$ & 0.81 & $95\pm17$ & No\\
   &            & $4.17\pm0.05$ & $2.7\pm0.3$, $2.0\pm0.2$\\
18 & 55608 & $4.07\pm0.04$  & $9\pm3$, $5.3\pm0.5$  & 96409-01-07-02 & $5.02\pm0.05$ & 0.87 & $138\pm7$ & No\\
   &            & $3.87\pm0.05$ & $3.5\pm0.3$, $2.8\pm0.3$\\
19 & 55610 & $3.9\pm0.1$  & $10\pm3$, $6.1\pm0.6$  & 96409-01-07-04 & $4.05\pm0.05$ & 0.89 & $80\pm18$ & No\\
   &           & $4.0\pm0.1$ & $4.6\pm0.4$, $4.1\pm0.4$\\
20 & 55618 & $2.54\pm0.04$  & $16\pm5$, $11\pm1$  & 96409-01-09-00 & $1.84\pm0.03$ & 0.88 & $14\pm15$ & No\\
   &           & $2.95\pm0.05$ & $9.5\pm0.9$, $8.6\pm0.8$\\
\hline
\end{tabular}
\end{table*}

 \section{The model}
 \label{sec:model}
 
 We use a semi-analytical, zonal jet model (see \citealt{mnw05,Maitra2009,Connors2017}). To calculate the dynamics, we assume the BHB launches a roughly isothermal jet that is accelerated to mildly relativistic velocities by internal pressure \citep{Crumley2017}.  We refer the reader to \cite{Connors2017} for the most up-to-date details and changes within the model prior to the changes discussed below, and to Table~\ref{tab:params} for a description of the key physical parameters of the model. We make two key improvements upon previous implementations of the model. \\
\indent Firstly, the calculation of IC emission within the jet has been improved to include multiple scattering events rather than adopting a single-scattering treatment. This allows the model to treat cases in which the jet-base is initially optically thick ($\tau \gg 1$), such that the flux contribution from higher IC scattering orders may be significant. In BHBs we expect the IC-emitting regions to remain optically thin \citep{Haardt1993,Done2007}. However, even as the IC region approaches $\tau \sim 1$, the higher-energy emission may become relevant since the Compton-y parameter (the number of scatterings $\times$ the energy shift per scattering) for a single electron goes as $y_i=16\Theta_i^2 \mathrm{Max}(\tau_i^2,\tau_i)$, where $\Theta_i \equiv kT_i/m_ec^2$ is the dimensionless energy of the electron, with $i$ representing a single electron within the full population. In simple terms, a large $y_i$ results in efficient scattering (more than one scattering order) but can be achieved in the following ways and will result in different spectral shapes: a IC spectrum with high $\tau$ ($\gg 1$) and moderate $\Theta_i$ ($\le 1$) will be a smooth power law, whereas at moderate $\tau$ ($\sim 1$) and high $\Theta_i$ ($\ge 1$) the IC spectrum will appear bumpy due to the separation of scattering orders (see, e.g., \citealt{Ghisellini2013}). Details of the multiple IC calculation used in this work can be found in a forthcoming paper (Ceccobello et al., in preparation). \\
 \indent Secondly, we have altered the jet height profile (z-profile) to improve the treatment of IC scattering in the first few zones of the jet. In all previous implementations of the model, a log scale is used between $z_{min}$ and $z_{max}$, where $z_{min} \sim 0.3~r_0$ and $z_{max}$ is a model parameter adjusted according to the source being modelled. Instead now we enforce $\Delta z = 2 r$ in all zones up to the cut-off of the Comptonising region (at $z_{cut}=100~r_0$), and space the remaining zones logarithmically up to $z_{max}$. In this way, we treat the input photon distribution for IC scattering as roughly isotropic without incurring any resolution-dependent errors, and without losing too much resolution in the effects of the jet profile at low heights. \\
\indent From here on we refer to this model as \texttt{agnjet}, thus maintaining consistency with its earlier applications to mildly-relativistic ($\gamma_{j}\sim~\mathrm{a~few}$) jets in Active Galactic Nuclei \citep{Markoff2015,Prieto2016, Connors2017,Crumley2017}.  
 
 \begin{table}
 \centering
  \caption{A list of the main input parameters of the \texttt{agnjet} model}
  \label{tab:params}
 \begin{tabular}{@{}cp{6cm}}
 \hline
 Parameter & Description \\
 \hline
 $N_{\rm j}$ ($L_{\rm Edd}$) & the normalised jet power.\\
 $r_0$ and $h_0$ ($r_{\rm g}$) & the radius and height (length) of the jet nozzle. The height is fixed at $h_0=2r_0$, such that the nozzle is a cylinder. \\
 $\Theta_{\rm e}$ ($kT_{\rm e}/mc^2$) & the electron temperature of the input distribution. \\
 $\beta_{\rm e}$ & the ratio of electron to magnetic energy density, ${U_{\rm e}}/{U_{\rm B}}$. \\
 $p$ & the power-law index of the accelerated electron distribution. \\
 $z_{\mathrm{acc}}$ ($r_g$) & the distance from the black hole along the jet axis where particle acceleration into a power-law distribution first begins.  \\
 $n_{\mathrm{nth}}$ & the fraction of particles accelerated at a distance $z_{\rm acc}$ from the black hole along the axis of the jet. \\
 $f_{\mathrm{sc}}$  & the scattering fraction, a measure of the efficiency with which electrons are accelerated at $z_{\mathrm{acc}}$, defined as ${\beta_{\mathrm{sh}}}^2/{\left(\lambda/R_{\mathrm{gyro}}\right)}$ where $\beta_{\mathrm{sh}}$ is the shock speed relative to the plasma, $\lambda$ is the scattering mean free path in the plasma at the shock region, and $R_{\mathrm{gyro}}$ is the gyroradius of the particles in the magnetic field. In reality we do not require a shock so this parameterisation can generally be seen as a measure of the acceleration efficiency, as it sets the maximum post-acceleration electron energy.\\
 \hline
 \end{tabular}
 \end{table}

 \section{Spectral fits}
 \label{sec:method}
 
 We perform all spectral fits in this work using the multiwavelength data analysis package \texttt{ISIS} \citep{Houck2000}, version 1.6.2-40. All models are forward-folded through the detector response matrices; when fitting to X-ray spectra this corresponds to the Proportional Counter Array (PCA) and High Energy X-ray Transmission Spectrometer (HEXTE) instrument responses, whereas data at all other wavelengths is assigned a "dummy" response equivalent to a detector of effective area = $1~\mathrm{m}^2$. Data at wavelengths outside the X-ray band are loaded into \texttt{ISIS} as flux measurements (shown in Table~\ref{tab:data}). We bin PCA spectra at a minimum signal-to-noise ratio of $S/N=4.5$, between energy limits of 3--45~keV or 3--20~keV depending on the availability of counts in the highest energy bins. A systematic error of 0.1\% is added to the PCA counts based on the improved calibration tool \texttt{PCACORR} \citep{Garcia_2014a}. We include HEXTE A/B spectra for the observations indicated in Table~\ref{tab:data}, and bin each at minimum signal-to-noise $S/N=4.5$ between energy limits 20--200~keV. At each stage of the fitting process we use the \texttt{ISIS} implementation of Markov Chain Monte Carlo (MCMC) parameter exploration routine \citep{Murphy2014}, based on the popular routine, \texttt{emcee}, developed by \cite{Foreman-Mackey2013}. In each case we initialize $50 \times n_{fp}$ walkers per free parameter, where $n_{fp}$ is the number of free parameters, and we run the MCMC chain until it has converged---we judge convergence as the point beyond which changes to the posterior probability distribution functions of the parameters are minimal, resulting in chains ranging in length between $10^3$--$10^4$ steps. 
 
 \subsection{X-ray spectral fits}
 \label{subsec:initialxrayfits}
 Before exploring broadband model fits to the quasi-simultaneous data of GX~339$-$4, we first fit phenomenological models to the available X-ray spectra in order to place prior constraints on nuisance parameters, allowing us to reduce the uncertainties in our broadband fits. These include the energy of the Gaussian iron emission line resulting from disc reflection, $E_{\mathrm{line}}$, and its corresponding line width, $\sigma_{line}$. We fix the interstellar Hydrogen column density to $n_\mathrm{H}=5 \times 10^{21}~\mathrm{cm^{-2}}$ based on previous X-ray spectral modelling of GX~339$-$4 \citep{Shidatsu2011,Garcia2015,Parker2016}, and on the cross-section adopted when correcting for extinction in the OIR \citep{Kong2002}. We consider 3 model classes, assigned according to the breadth of X-ray band coverage and the number of X-ray counts in the spectra: 
 
 \begin{itemize}
 \item X1: \texttt{tbabs$\times$[powerlaw+gaussian]}
 \item X2: \texttt{tbabs$\times$[reflect(powerlaw)+gaussian]}
 \item X3: \texttt{tbabs$\times$[reflect(powerlaw$\times$highecut)+gaussian]}. 
 \end{itemize}
 
\begin{figure}
 \includegraphics[width=1.1\linewidth]{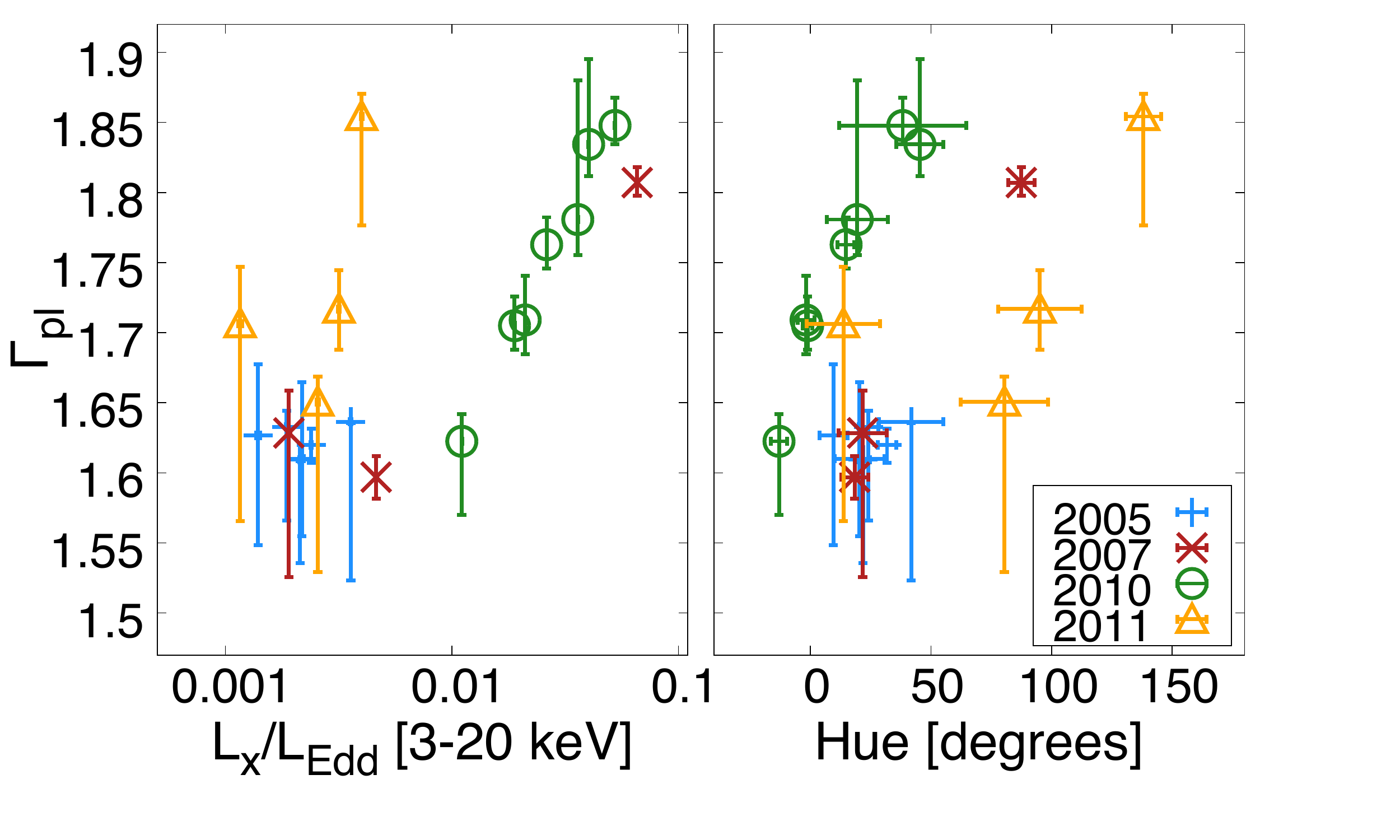}\vspace{-0.5cm}
 \caption{The power-law spectral index ($\Gamma_{\mathrm{pl}}$, derived from initial spectral fits to all 20 X-ray spectra) against the unfolded data luminosity (left) between 3--20 keV and the power-spectral hue (right). The key shows how the data are divided by observation year.}
 \label{fig:gamma_trend}
 \end{figure}

\begin{figure}
 \centering
 \includegraphics[width=\linewidth]{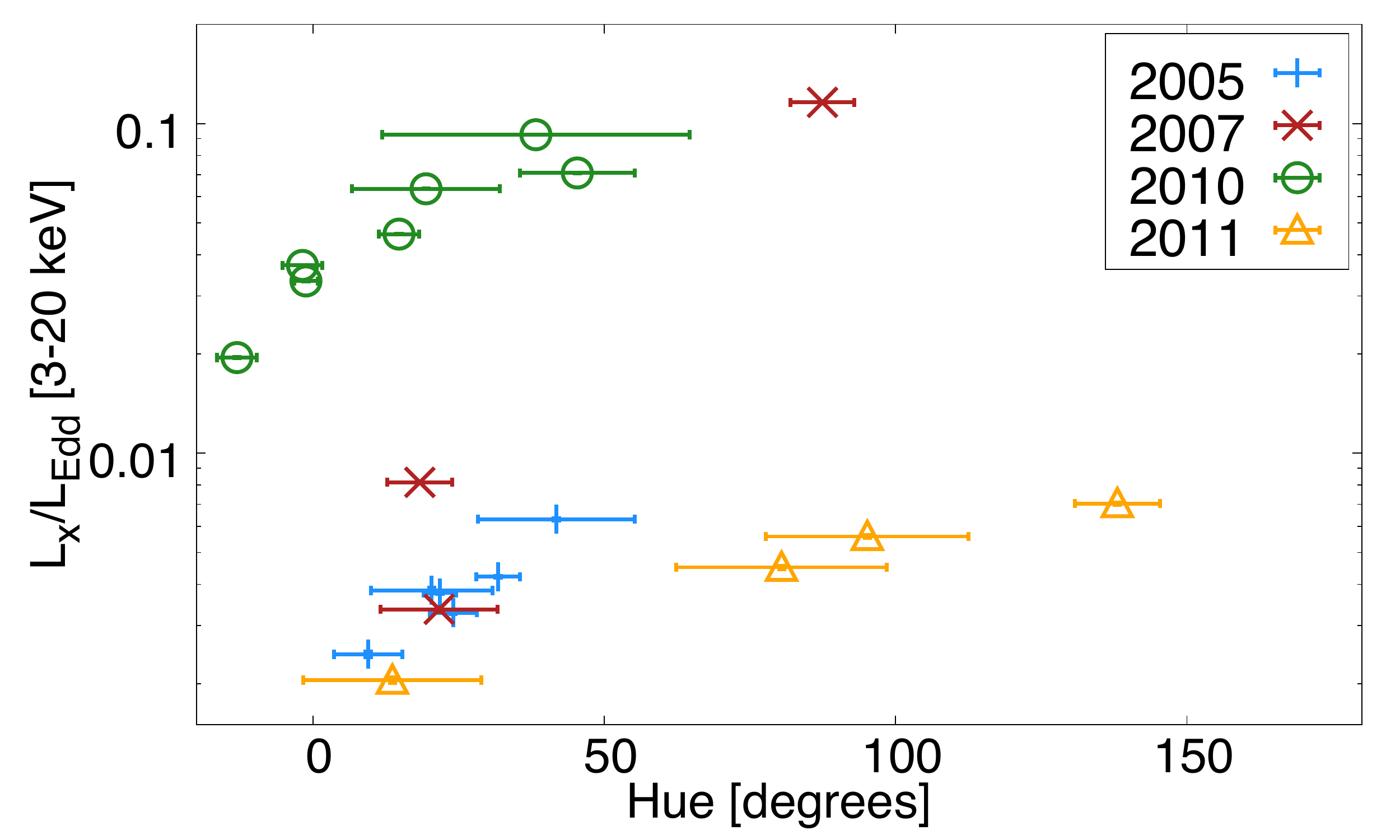}\vspace{-0.5cm}
 \caption{The Eddington-scaled X-ray luminosity of all X-ray spectra against the power-spectral hue derived from the light curves. The key shows how the data are divided by observation year. }
 \label{fig:hue_v_lx}
 \end{figure}

 The reflection convolution model \texttt{reflect} is that of \cite{Magdziarz1995}, and we adopt this in preference to more recent reflection models (\texttt{RELXILL}; \citealt{Dauser2014,Garcia2014}, \texttt{REFLIONX}; \citealt{Ross1999,Ross2005}) since it convolves an arbitrary input spectrum, whereas the more recent models rely on robust model tables that are expensive to produce. The absorption model \texttt{tbabs} is described in \cite{Wilms2000}. We adopt the solar abundances of \cite{Wilms2000} and set the photo-ionisation cross-sections according to \cite{Verner1996}.  Model~X1 is most likely to provide a sufficient fit to those X-ray spectra with low source counts, Model~X2 (a reflected power law) will apply when source counts are high enough to distinguish a break in the spectrum at $E \sim 10~\mathrm{keV}$, characteristic of a reflected X-ray spectrum, and Model~X3 applies to only one spectrum for which we see a clear visible cutoff in the spectrum. We perform Markov Chain Monte Carlo (MCMC) parameter exploration on each X-ray spectral fit in order to characterise the posterior probability distribution functions (PDF) of $E_{\mathrm{line}}$ and $\sigma_{line}$. We fix $E_{\mathrm{line}}$ and $\sigma_{line}$ based on these fits, and carry those values forward to our broadband spectral modelling described in Section \ref{subsec:bbmodelling}. \\
 \indent Figure~\ref{fig:gamma_trend} shows the evolution of $\Gamma$, the power-law spectral index, against both the power-spectral hue and the unfolded data luminosity (assuming $D=8~\mathrm{kpc}$). There exists a clear dichotomy between the more luminous X-ray spectra of the 2010 outburst rise and 2007 single observation of its outburst, with the observations in the decay phases of 2005, 2007, and 2011. The spectrum appears to slightly soften with increasing luminosity/hue. The increase in power-spectral hue is coincident with increasing luminosity as the source evolves through its outburst, but with two distinct trends depending on whether the source is in the rise or decay of an outburst, making the plot appear like the mirror of the hardness-intensity diagram (see Figure~\ref{fig:hue_v_lx}). Many previous works find $\Gamma$ to be mostly constant during the rising hard state of GX~339$-$4 \citep{Wilms1999,Zdziarski2004,Plant2014,Garcia2015}, so the fact that we see a slight positive correlation may be related to the model treatment, in particular the reflection model used, as well as the treatment of the data. For example, \cite{Garcia2015} combine spectra across ranges of X-ray hardness, and use a different model for the X-ray reflection, which likely leads to contrasting photon indices. However, we note such a positive trend does agree with the broader trends seen in multiple BHBs (see e.g. \citealt{RemMc2006}), and is coincident with the narrowing and strengthening of broadband X-ray variability \citep{Heil2015}, and brighter radio jets \citep{Fender2006}.
 
\subsection{Broadband spectral modelling}
 \label{subsec:bbmodelling}
Next, we fit all 20 of the quasi-simultaneous broadband spectra energy distributions (SEDs) of GX~339$-$4 with two more physically motivated models, with the goal of tracking the trends in the physical parameters of the jet and corona or inner accretion flow. We find that the X-ray spectra are best fit by a coronal-like IC-scattering plasma, in which the scattering electrons are at temperatures of $kT_\mathrm{e}\sim$ hundreds of keV, and the plasma has optical depths in the range 0.1--1. We also find that the hotter jet electrons likely contribute a non-negligible flux in the X-ray band as a result of SSC.\\
\indent The two models we adopt are as follows:  

\begin{itemize}

\item B1: an absorbed, reflected jet component + Gaussian line: \texttt{tbabs}$\times$[{\texttt{reflect}\texttt{(agnjet)+gaussian}}]\item B2: the sum of absorbed, reflected jet and coronal components + Gaussian line:\\ \texttt{tbabs}$\times$[\texttt{reflect}\texttt{(agnjet}\texttt{+nthcomp})\texttt{+gaussian}].

\end{itemize}

\begin{figure}
\includegraphics[width=0.4\textwidth,angle=270]{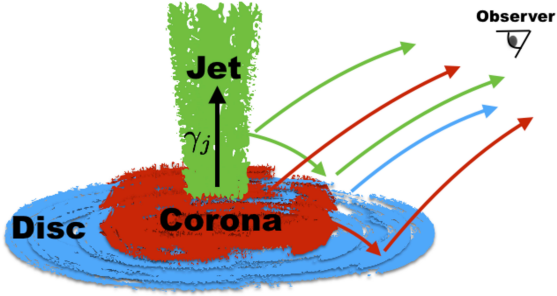}
\vspace{-1.2cm}
\caption{Diagram of a corona + jet model for a BHB, representing spectral Model~B2. The thin disc is truncated to radii on the order of 2--10~$r_{\rm g}$, and an optically thin compact corona exists within the inner accretion flow, with electron temperatures $kT_{e}\sim$ hundreds of keV. The jet electrons are relativistic, with $kT_e \ge 511~\mathrm{keV}$, and the plasma has bulk motion in the z-direction with $\gamma_j \sim$ on the order of 1 to a few. The observer sees emission from the jet in the form of synchrotron, SSC, and IC scattering of disc photons, as well as IC emission from the corona, and blackbody emission from the disc. The jet and coronal X-ray components irradiate the disc, resulting in a reflected X-ray spectrum.}
\label{fig:corona_jet}
\end{figure}

Here \texttt{agnjet} represents the jet and outer standard disc components, and \texttt{nthcomp} represents a spherical corona in the inner regions of the accretion flow \citep{Zdziarski1996,Zycki1999}, and thus B2 is only distinguishable from B1 through the additional coronal thermal Comtponisation component. Figure~\ref{fig:corona_jet} shows a diagram of the setup which represents spectral Model~B2. Whilst \texttt{agnjet} does in fact include a coronal-like jet base \citep{mnw05}, its treatment of SSC is purely relativistic, allowing only for photon-scattering electrons at $\Theta_{\rm e}\equiv{kT_e/m_ec^2}\ge1$ ($kT_e\ge511~\mathrm{keV}$)---this is due to the expectation that energy is dissipated to the electrons quite rapidly within the jet, giving rise to high synchrotron fluxes in the radio bands in regions further out along its axis; conservation arguments suggest similarly hot electrons ($\Theta_{\rm e}>1$) at the base of the jet (see e.g. \citealt{mnw05}). Popular models for the X-ray spectra observed in BHB hard states typically include \texttt{nthcomp} in which a thermal population of electrons at roughly $\Theta_{\rm e} \sim 0.02$--$0.2$ IC scatter the soft blackbody component of the accretion flow with seed photons temperatures in the range $kT_{\mathrm{BB}}\sim0.01$--$1~\mathrm{keV}$, set by the inner disc temperature (see e.g. \citealt{Haardt1993,Done2007} and references therein). Thus in our model-fitting treatment we choose to test a combination of both emission components in order to determine the relevant importance of each during an evolving outburst.

\begin{figure}
\includegraphics[width=\linewidth]{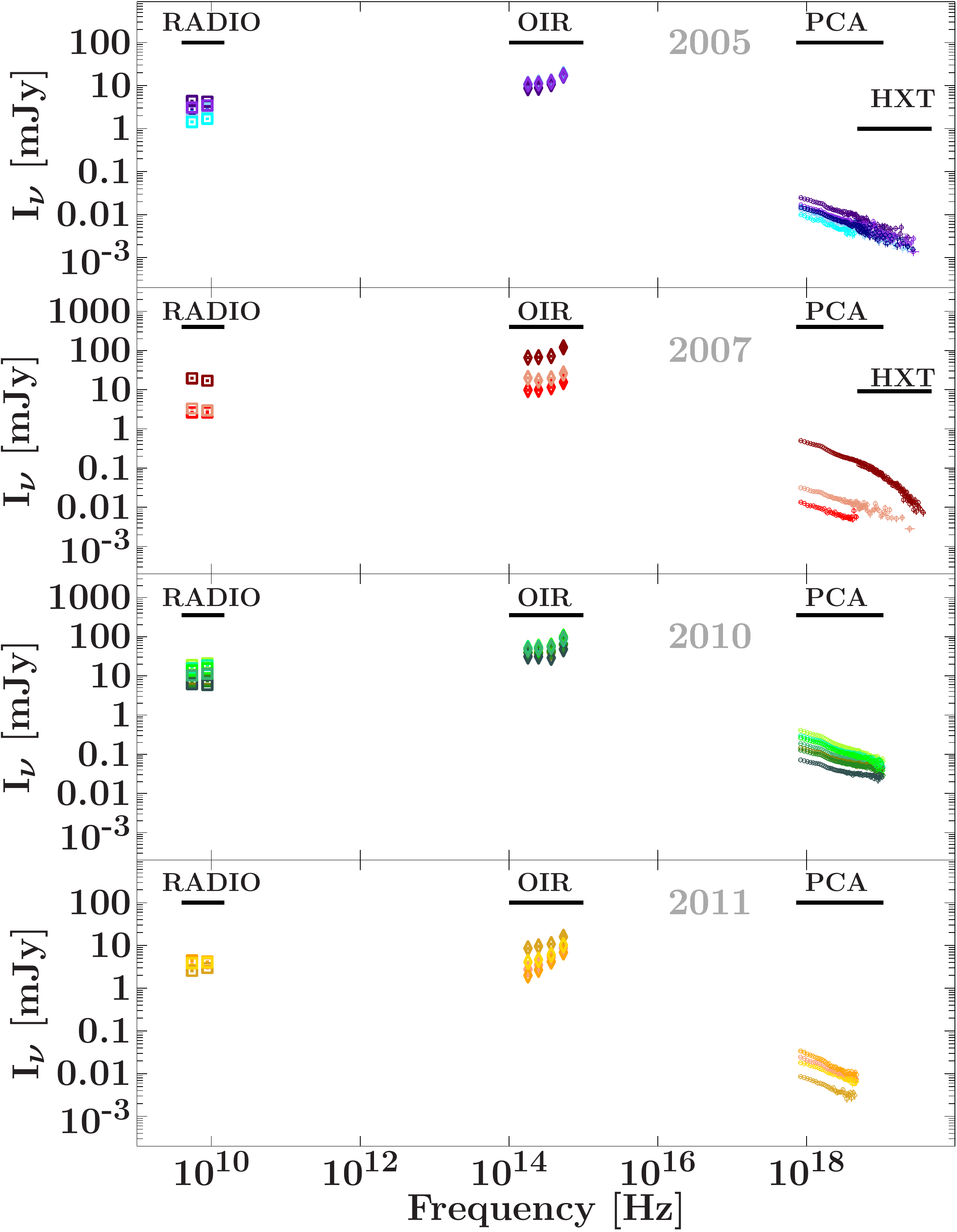}\vspace{-0.5cm}
\caption{All 20 broadband spectra split into panels based on observation year. The observed flux density is shown as a function of frequency, with the radio, OIR, and X-ray bands indicated. Unfolded fluxes are calculated independently of the spectral model.}
\label{fig:all_data}
\end{figure}

 \begin{figure*}
 \includegraphics[width=\textwidth]{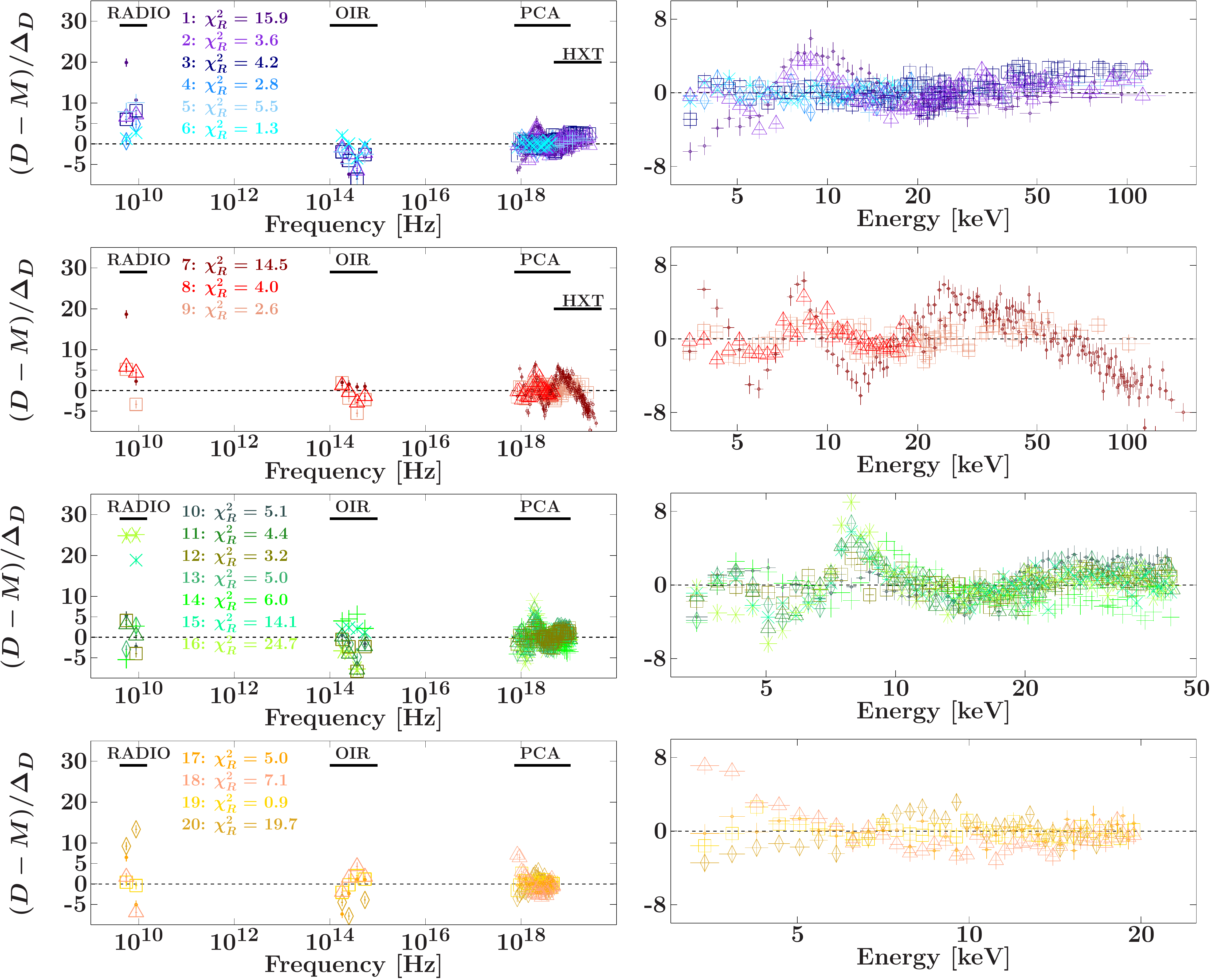}
 \caption{Standardised residuals ((data - model) / uncertainty) of fits to all 20 broadband spectra of GX~339$-$4 with Model~B1 (jet IC-dominated X-ray spectra): \texttt{tbabs $\times$ [reflect(agnjet)+gaussian]}, with typical $\chi^2_{R}\sim$ a few to 10s. Each panel shows fits to observations within 2005, 2007, 2010, and 2011 respectively. Radio, OIR, and X-ray data are labelled, and different symbols/colours indicate the broadband data-model residuals for each fit. The X-ray spectra are dominated by jet SSC and IC scattering of disc photons in the jet, as well as reflection off the disc.}
 \label{fig:allfits_ssc}
 \end{figure*}
 
  \begin{figure*}
 \includegraphics[width=\textwidth]{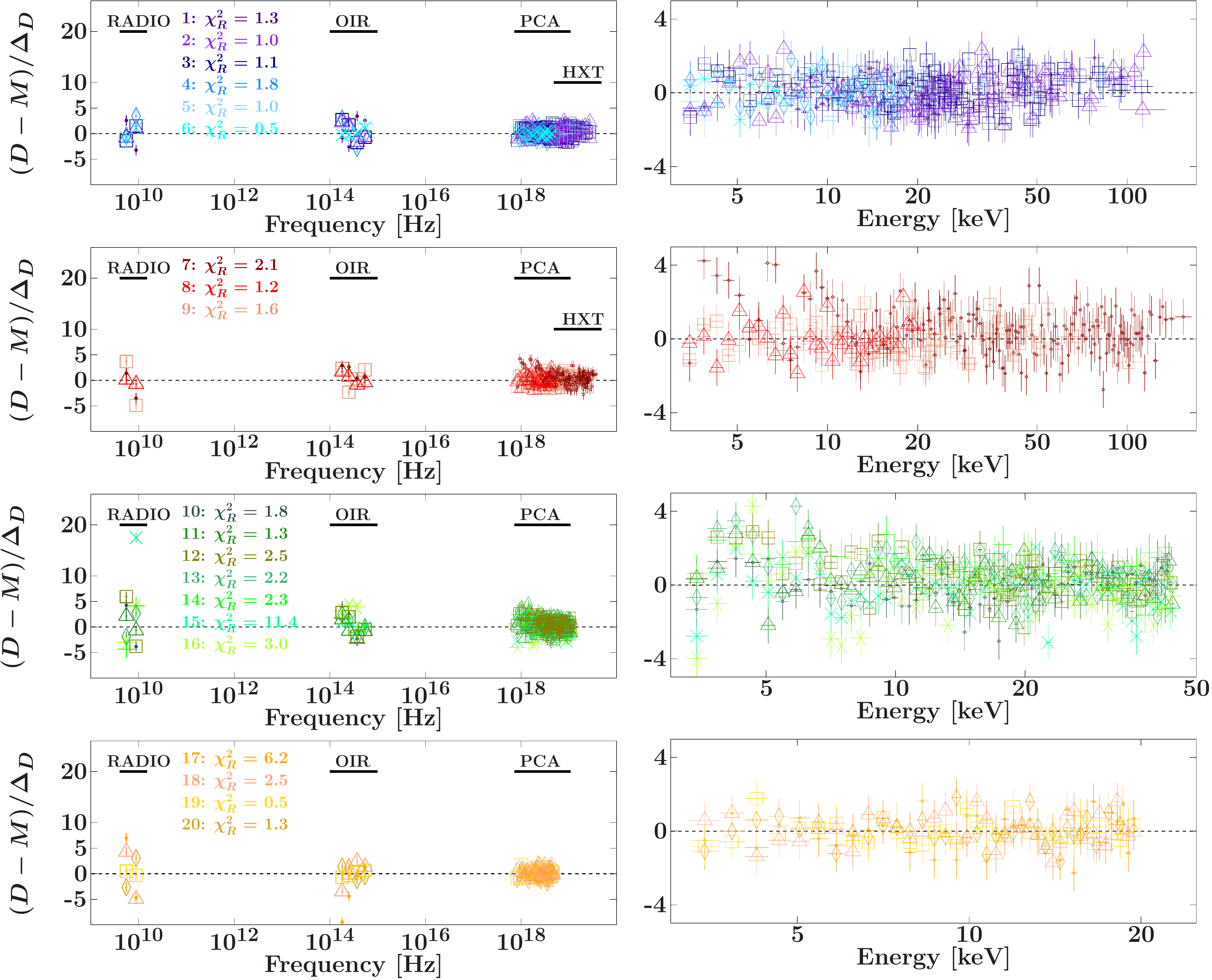}
  \caption{Standardised residuals ((data - model) / uncertainty) of fits to all 20 broadband spectra of GX~339$-$4 with Model~B2 (coronal IC-dominated X-ray spectra): \texttt{tbabs $\times$ [reflect(agnjet+nthcomp)+gaussian]}, with typical $\chi^2_{R}\sim$1--2, with some outliers. Each panel shows fits to observations within 2005, 2007, 2010, and 2011 respectively. Radio, OIR, and X-ray data are labelled, and different symbols/colours indicate the broadband data-model residuals for each fit. The coronal emission (\texttt{nthcomp}) dominates the X-ray spectra, with SSC and IC scattering of disc photons in the jet contributing, and all emission reflecting off the disc. }
 \label{fig:allfits_nthcomp}
 \end{figure*}

 \begin{figure*}
 \includegraphics[width=0.47\textwidth]{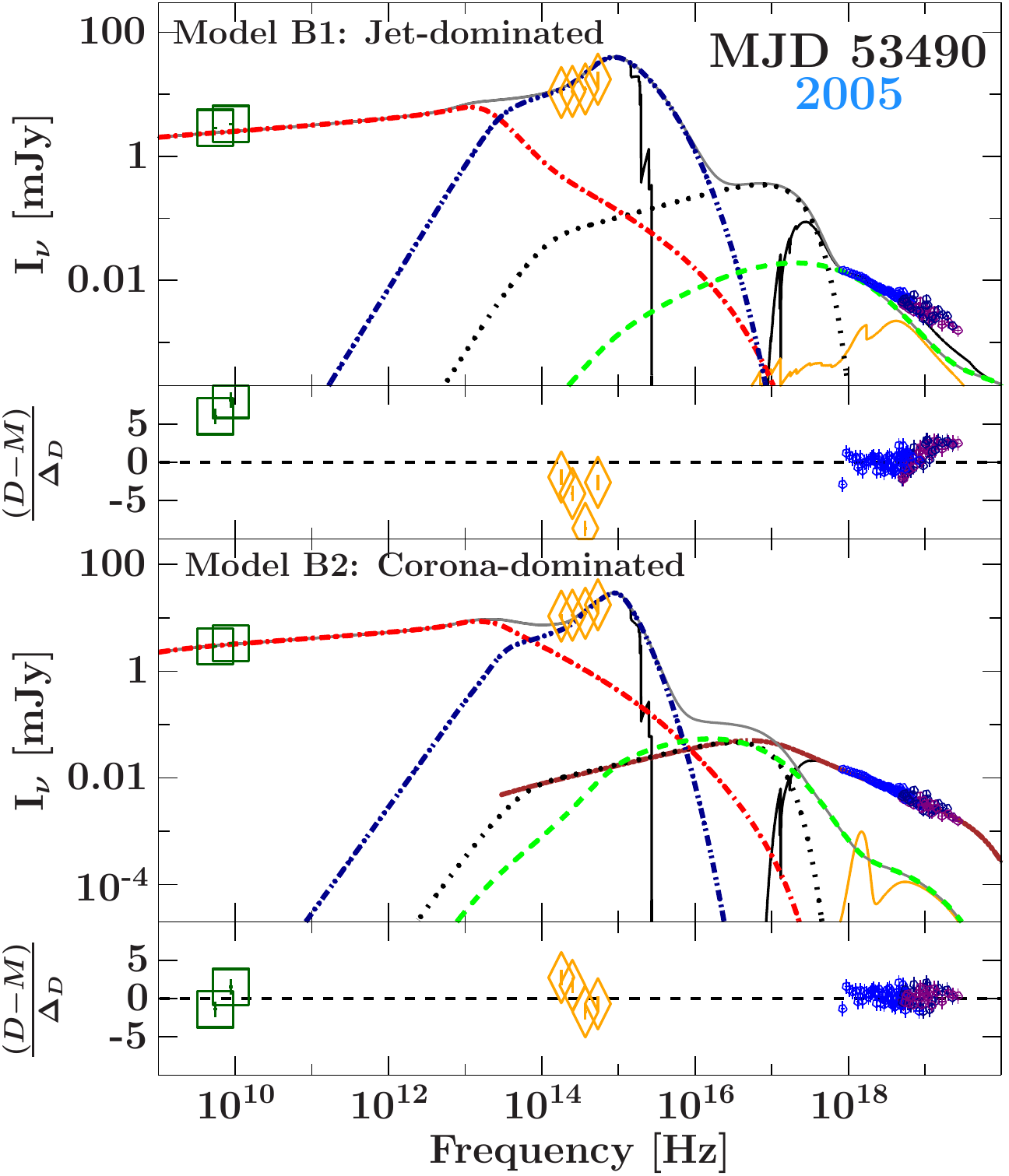}
  \includegraphics[width=0.47\textwidth]{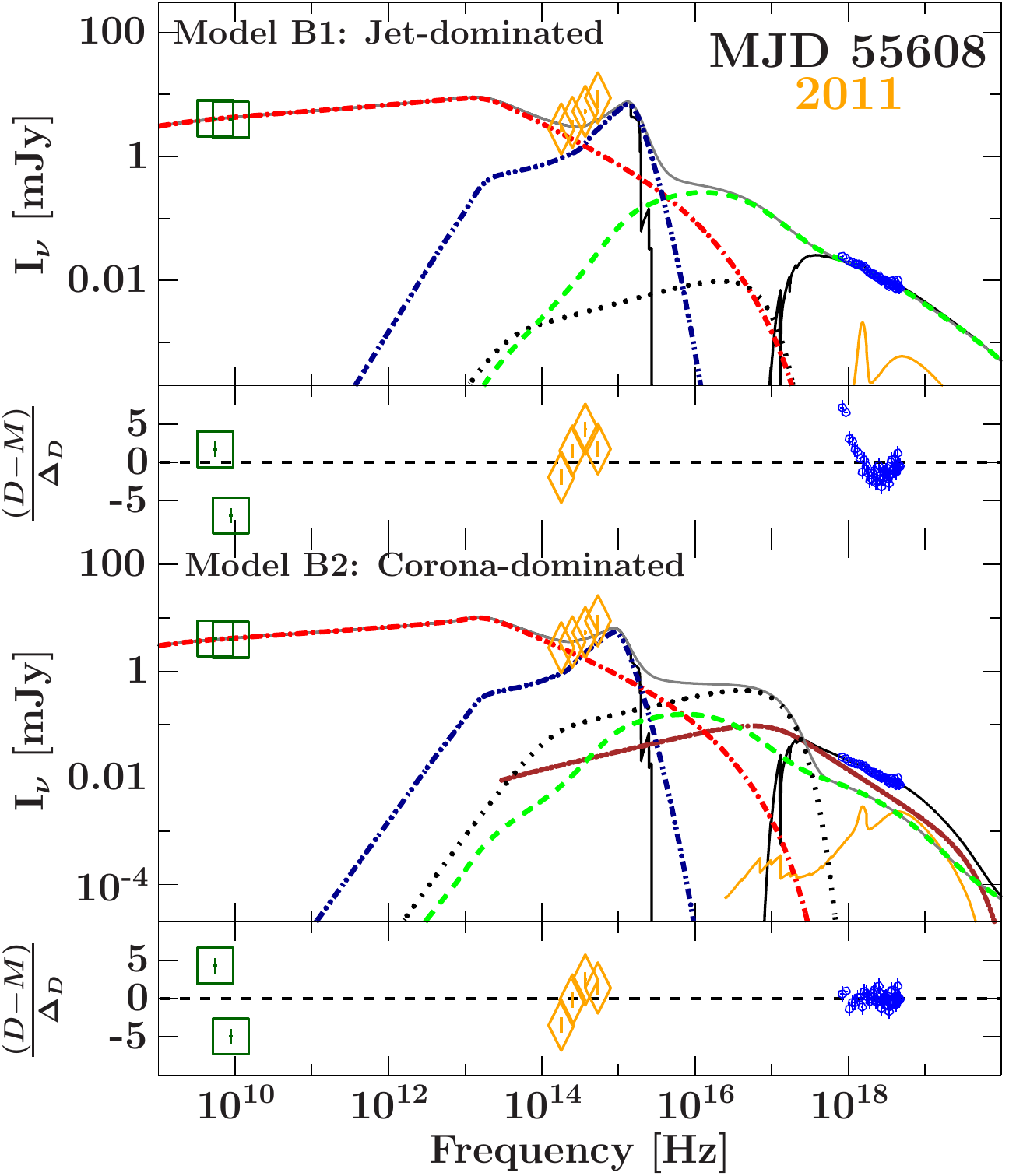}
    \includegraphics[width=0.47\textwidth]{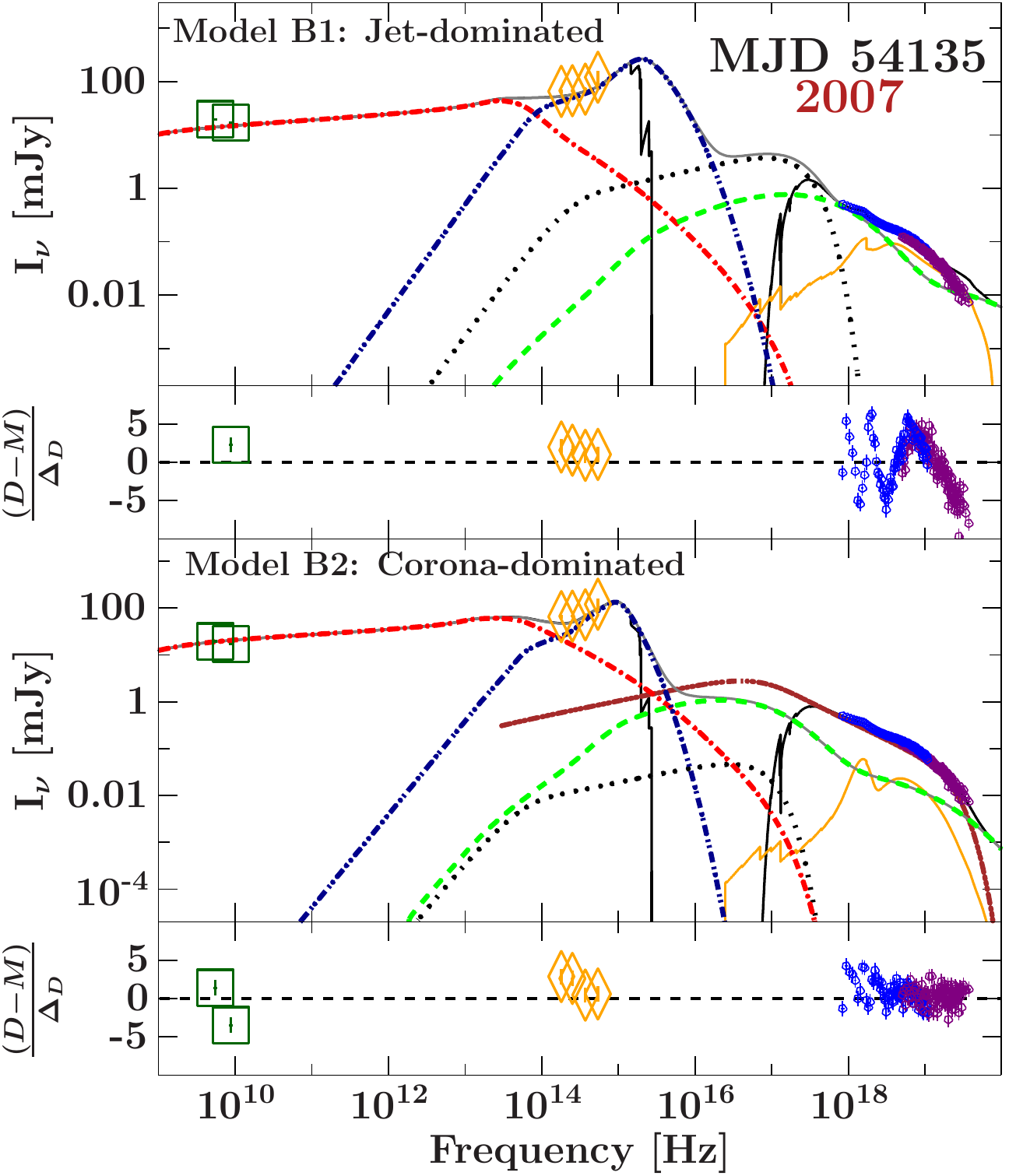}
  \includegraphics[width=0.47\textwidth]{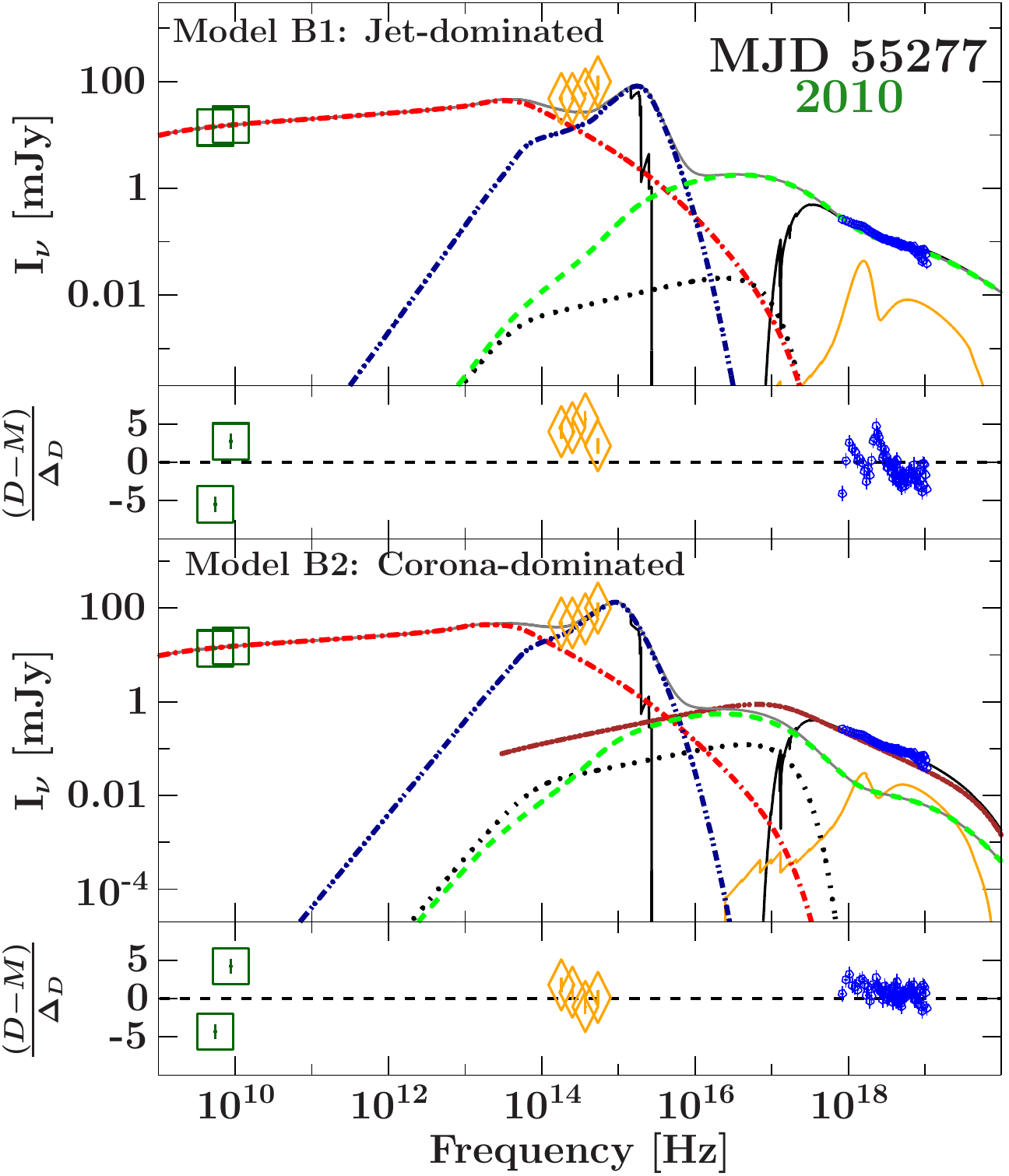}
 \caption{An expanded view of the fit to broadband spectra 3 (top left), 18 (top right), 8 (bottom left) and 14 (bottom right) from GX~339$-$4 outburst decays in years 2005 and 2011, and outburst rises in years 2007 and 2010 respectively (see Table~\ref{tab:conf}). The top panels show the full broadband fit with Model~B1 (jet IC-dominated X-ray spectrum): \texttt{tbabs $\times$ [reflect(agnjet)+gaussian]}. The bottom panels show the same spectrum fit with Model~B2 (coronal IC-dominated X-ray spectrum): \texttt{tbabs $\times$ [reflect(agnjet+nthcomp)+gaussian]}. Radio data are marked with  green squares, OIR data with orange triangles, and \textit{RXTE}-PCA, HXT A, and HXT B with blue, purple and dark blue circles respectively. The total broadband jet spectrum is shown with solid gray lines. The individual jet spectral components shown are SSC/IC (green dashed lines), pre-acceleration thermal synchrotron (blue dot-dot-dashed lines), post-acceleration synchrotron (red dot-dashed lines), and the accretion disc blackbody spectrum (black dotted line). The reflection component (Gaussian iron line included) is shown in orange, and the coronal component in the right-hand panels is shown as a brown triple-dot-dashed line. The solid black line shows the total absorbed model spectrum. The disc component of \texttt{agnjet} and coronal component of \texttt{nthcomp} normalisations are treated separately. In the panels beneath each fit we show the standardised $\chi$-residuals, (data-model)/uncertainty.}
 \label{fig:specfits}
 \end{figure*}

\begin{figure}
 \centering
 \includegraphics[width=\linewidth]{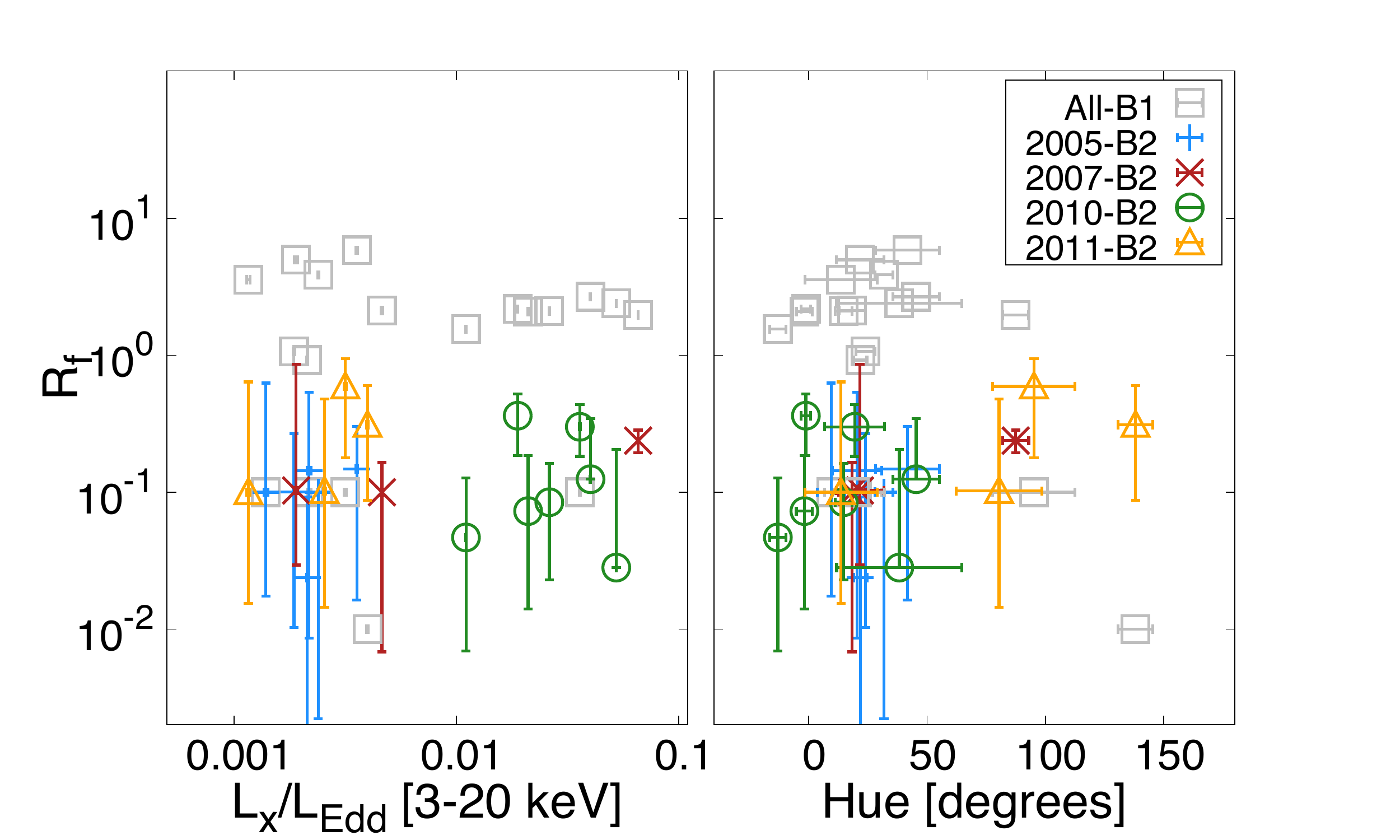}
 \caption{The maximum likelihood estimates of the reflection fraction, $R_{\mathrm{f}}$, as a function of Eddington-scaled X-ray data luminosity (left) and power-spectral hue (right). Hollow gray squares show the parameters derived from fits of Model~B1, and other symbols (indicated in the key) show fits of Model~B2.}
 \label{fig:Rf}
 \end{figure}

\begin{figure}
 \centering
 \includegraphics[width=1.1\linewidth]{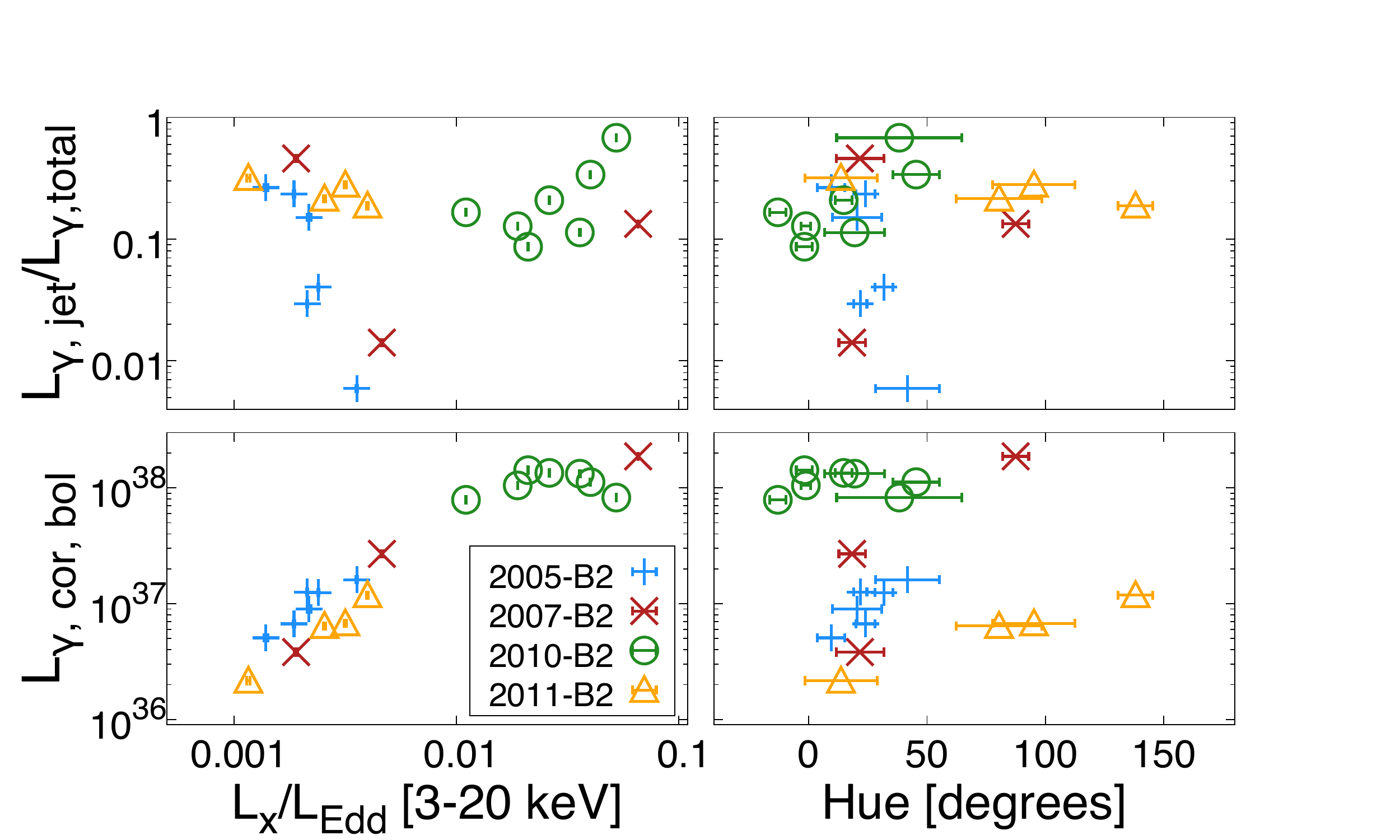}\vspace{-0.3cm}
 \caption{The integrated radiative jet luminosity in the 3--100~keV X-ray band ($L_{\rm \gamma, jet}$) as a fraction of the total 3--100~keV model luminosity ($L_{\rm \gamma, total}$) (top), and the Bolometric coronal luminosity in ergs/s (bottom), shown as a function of Eddington-scaled X-ray data luminosity (left) and power-spectral hue (right). Luminosities are shown only for fits of Model~B2, divided by observation year. Uncertainties are shown only for $L_{\rm X}$ and hue. The jet radiative luminosity has not been corrected for Doppler beaming. }
 \label{fig:Ljet}
 \end{figure}

\indent We fix the location where particle acceleration starts to $\log_{10} [z_{\rm acc}] = 3.5$ since this is the approximate location at which a non-thermal population of electrons is generated in GX~339$-$4, according to the location of the variable self-absorption spectral break \citep{Markoff2003,Gandhi2011}. If we interpret a time lag of 100~ms \citep{Kalamkar2016} as being caused by the delay of plasma flow through the jet, this would imply a distance scale of $z \ge 0.1s \times \gamma_{\rm j}\beta_{\rm j} c \sim 10^3~r_{\rm g}$, where $\gamma_{\rm j}$ is the jet bulk Lorentz factor, and $\beta_{\rm j}$ is the jet bulk velocity. This distance is conservative given that the jet is assumed to travel at constant velocity---the jet could accelerate efficiently along its axis, as is the case with \texttt{agnjet}. It is preferable to keep $z_{\rm acc}$ fixed at this value since the data coverage provides limited constraints on its value, and the self-absorption break is variable on timescales shorter than 24 hours (e.g., \citealt{Gandhi2010}, and see Section~\ref{subsubsec:oir}). We fix the fraction of particles accelerated at $z_{\mathrm{acc}}$ to $n_{\rm nth}=0.1$, based on current studies of particle acceleration across mildly-relativistic shocks (e.g. \citealt{ss11}). We also set the power law index of the accelerated electrons to $p=2.2$, in accordance with typical values expected for the Fermi diffusive shock acceleration process (e.g., \citealt{Blandford1978,Drury1983}). Whilst it is desirable to allow $p$ to vary freely, since it influences other key model parameters (due to the energy supplied to the outer post-acceleration regions of the jet by the higher energy particles), it is not easily contrained due to the lack of data between the OIR and X-ray regions. We fix the scattering fraction (which determines the maximum energy to which particle are accelerated, thus setting an upper bound on the power law synchrotron cutoff) to $f_{\rm sc}=10^{-6}$ to ensure no significant direct contribution of optically thin synchrotron to the X-ray spectrum. This choice to suppress the X-ray synchrotron contribution is motivated by our objective to constrain the jet IC contribution to the X-ray spectrum of GX~339$-$4, and to limit degeneracies in tracking the jet properties in outburst. Also, a dominant jet synchrotron component in the X-rays likely predicts hard X-ray lags on timescales far shorter than those observed in GX~339$-$4 \citep{Nowak1999,Belloni2005,Altamirano2015}, based on short expected particle acceleration timescales within the jet \citep{Connors2017}. We note, on the other hand, that a non-negligible contribution in the X-ray band from synchrotron photons may be present without violating the observed lags. 

 The fundamental parameters of interest in \texttt{agnjet} are the normalised jet power, $N_{\rm j}$, the radius of the jet base, $r_0$, the dimensionless initial electron temperature, $\Theta_{\rm e}$, and the ratio of energy density between the electrons and magnetic field at the jet base, $\beta_{\rm e}$, all of which remain free parameters in the minimisation process.  
 
 We set the input photon distribution of \texttt{nthcomp} as a multi-temperature disc blackbody, and tie the disc temperature $T_{\mathrm{BB}}$ to the multi-temperature disc component within \texttt{agnjet}, $T_{\rm in}$, and allow the inner disc radius, $R_{\rm in}$, to vary between $\sim1.5$--10$r_{\rm g}$. Values higher than $10r_{\rm g}$ for the typical disc temperatures result in unrealistically high accretion rates, close to the Eddington rate for the given black hole mass, and lead to excessive soft X-ray fluxes which plainly disagree with the observations. The coronal electron temperature, $kT_{\mathrm{e,cor}}$, and spectral index, $\Gamma_{\mathrm{cor}}$ are free parameters of the model. The disc and coronal normalisations are treated separately, the disc being normalised by the black hole mass and distance parameters of \texttt{agnjet}, and coronal normalisation an independent counts normalisation constant inherent to the \texttt{nthcomp} model. As discussed in Section \ref{subsec:initialxrayfits}, we fix the centroid energy and width of the Gaussian iron line, $E_{\mathrm{line}}$ and $\sigma_\mathrm{line}$, according to the initial fits to each individual X-ray spectrum, after having fully explored the parameter distributions using MCMC parameter exploration. 
 
 After using minimization to converge as closely as possible on the global minimum of the fits to all 20 broadband spectra (characterized by the $\chi^2$ fit statistic), we initialise MCMC walkers around the maximum likelihood estimates of parameters in Model~B2, allowing the parameter search to explore the contributions of \texttt{agnjet} and \texttt{nthcomp}. Each MCMC chain is allowed to run for $10^3$--$10^4$ steps (on the basis of computational time constraints) such that the resultant posterior PDFs of the model parameters show coverage of the broad range intrinsic to models B1 and B2, and there is no longer significant evolution in those PDFs. We then discard the first 80\% of the MCMC chains, as this is well beyond the characteristic "burn-in" phase after which the chain is close to convergence, and the resultant walker distribution is populated enough to cover the parameter space. 
 
 Figure~\ref{fig:all_data} shows the full broadband spectrum from each observation, in unfolded flux space, colour-coded to show the evolution of each outburst in 4 separate panels, which we include to give clarity on the spectral evolution of GX~339$-$4 associated with all the datasets we model here.
 
 Figures~\ref{fig:allfits_ssc} and \ref{fig:allfits_nthcomp} show the standardised data$-$model residuals from model fits to all 20 broadband spectra of GX~339$-$4. Specifically, Figure~\ref{fig:allfits_ssc} shows residuals in which IC emission from \texttt{agnjet} is the dominant X-ray spectral component, i.e., Model~B1. Figure~\ref{fig:allfits_nthcomp} shows residuals from fits to the same spectra in which the coronal IC emission of \texttt{nthcomp} dominates the X-ray spectra, i.e., Model~B2. The first thing we notice, is that Model~B2 (due to the additional presence of a corona, i.e., \texttt{nthcomp}) provides a much better fit to the X-ray spectra than Model~B1 in each case ($\chi^2_{R}\sim1$--2 for Model~B2 compared to $\chi^2_{R}\ge$ a few for Model~B1), due to the lower electron temperatures and higher optical depths inherent to the model: $kT_{\mathrm{e,cor}}/mc^2 \sim$ 0.02--0.4 (though we note that solutions permit coronal electron temperatures of up to $\sim1000$~keV, or $\Theta_{\rm e}\sim2$, see Table~\ref{tab:conf}). This is seen more clearly in Figure~\ref{fig:specfits} where fits to four spectra (one from each year of observation) are shown in more detail, emphasizing the model components. One can see that constraints provided by the radio and OIR data result in a non-negligible contribution from the jet in the X-ray band. 

We find optical depths in the corona from all our B2 fits to be in the range $\tau \sim 0.1$--$1$, assuming the corona has a spherical geometry, and this is the key discriminator between the jet and coronal models we have considered. The electrons at the base of the jet in \texttt{agnjet} are strictly relativistic, and they remain quasi-isothermal throughout the jet. The optical depth in the jet base ranges between $\tau \sim10^{-4}$--$10^{-2}$. These conditions give rise to an emergent IC spectrum that is not a power law, but instead has significant curvature. This spectral curvature is a distinguishing feature of thermal Comptonisation (including SSC from a thermal particle distribution) with low optical depth ($\tau \ll 1$) and high electron temperature ($\Theta_{\rm e} \ge 1$). Due to the spectral curvature of the IC emission in \texttt{agnjet} in Model~B1, the reflection fraction ($R_{\mathrm{f}}$) systematically increases (see Figures~\ref{fig:specfits} and \ref{fig:Rf}). Thus fits with Model~B1, in which the jet IC emission dominates, require much stronger reflection than fits with Model~B2 in which the coronal IC emission dominates (see Figure~\ref{fig:Rf}). This increase in $R_{\mathrm{f}}$ is in disagreement with values derived from simpler X-ray spectral fits, and it is unlikely that a curved IC spectrum from the jet conspires with reflection to reproduce stable power law spectra over time. There is also a prominent apparent residual feature around 8--9~keV in most of the Model~B1 fits, which is clearly visible the residuals of all plots shown in Figure~\ref{fig:allfits_ssc}. This is a consequence of the curvature of the SSC component along with the dominance of the reflection component. In contrast, in fits of Model~B2, the power-law-like continuum of \texttt{nthcomp} fits well to the spectrum with the need for only a minimal reflection component and a Gaussian line to account for the line and added curvature. 

We also notice in fits of Model~B2 (see Figure~\ref{fig:specfits}) that the presence of non-negligible IC emission from the jet (we find that the jet contributes a range of a few up to $\sim50$\% of the continuum flux in the 3--100~keV band) acts to skew the shape of the model coronal spectrum. The extent to which the jet contributes to the X-ray spectrum is illustrated quantitatively for all the fits in Figure~\ref{fig:Ljet}, which shows the 3--100~keV jet radiative luminosity $L_{\rm \gamma, jet}$ with respect to the total radiative luminosity of the model $L_{\rm \gamma, total}$, as a function of $L_{\rm X}$ and hue. The Bolometric coronal luminosity is also shown, which highlights the progressive brightening of the corona (ranigng from $<1\%$--$10\%$~$L_{\rm Edd}$) with $L_{\rm X}$ for clarity. We see no obvious trends in $L_{\rm \gamma, jet}$, but it is noticeable that the jet can have a significant contribution in the X-ray, and this is due to the jet dominating the OIR and radio fluxes. If the jet contributes to the X-ray spectrum, the corona may either have a softer or harder spectral shape than would be concluded if the jet were to be ignored. This possibility then opens up a myriad of interesting questions to explore regarding the contributions of each of these components to the X-ray variability of BHBs, in particular the hard X-ray lags (e.g., \citealt{Nowak1999,Belloni2005,Altamirano2015}).

 The key results of our broadband model fits are that a coronal-like IC-scattering spectrum fits best to the data, whereby the electrons doing the scattering are at hundreds of keV, the plasma has optical depths in the range 0.1--1, and the photons being scattering originate in the disc with temperature around 0.1--1~keV. Such a scattering plasma produces the canonical power-law in the X-ray, with reflection reproducing the iron emission line and Compton reflection hump. However, SSC emission from hotter jet electrons within a plasma of optical depth in the range $\sim10^{-4}$--$10^{-2}$ likely has a non-negligible contribution in the X-ray, and this is constrained by the radio and OIR data, which can be well modelled by synchrotron from the hot jet electrons. Again we stress that any contribution of jet synchrotron emission to the X-ray bands has been suppressed, and a modelling treatment that includes that optically thin synchrotron component would add futher nuance to this conclusion.

\subsection{Jet parameter trends}
\label{subsec:global_trends}
We explore trends in the physical properties of the jet as a function of both the Eddington-scaled X-ray luminosity and variability properties (gauged by the power-spectral hue) of GX~339$-$4 during the different stages of its outburst rise and decay. Even though the coronal IC component dominates the X-ray spectra in Model~B2 (and Model~B2 provides superior fits to all our broadband datasets than Model~B1), the trends in key physical jet parameters (such as jet power) are similar to those found when fitting with Model~B1. This is because the radio and OIR data allow constraints on the jet physics. The main differences between the two models are firstly that the SSC emission from the jet is necessarily suppressed in Model~B2 to accommodate the dominant coronal component in the X-ray, and secondly that the reflection features are less prominent in Model~B2 (since the power-law like coronal spectrum accounts for most of the fit residuals in the X-ray). 

\begin{figure*}
 \centering
 \includegraphics[width=0.48\linewidth]{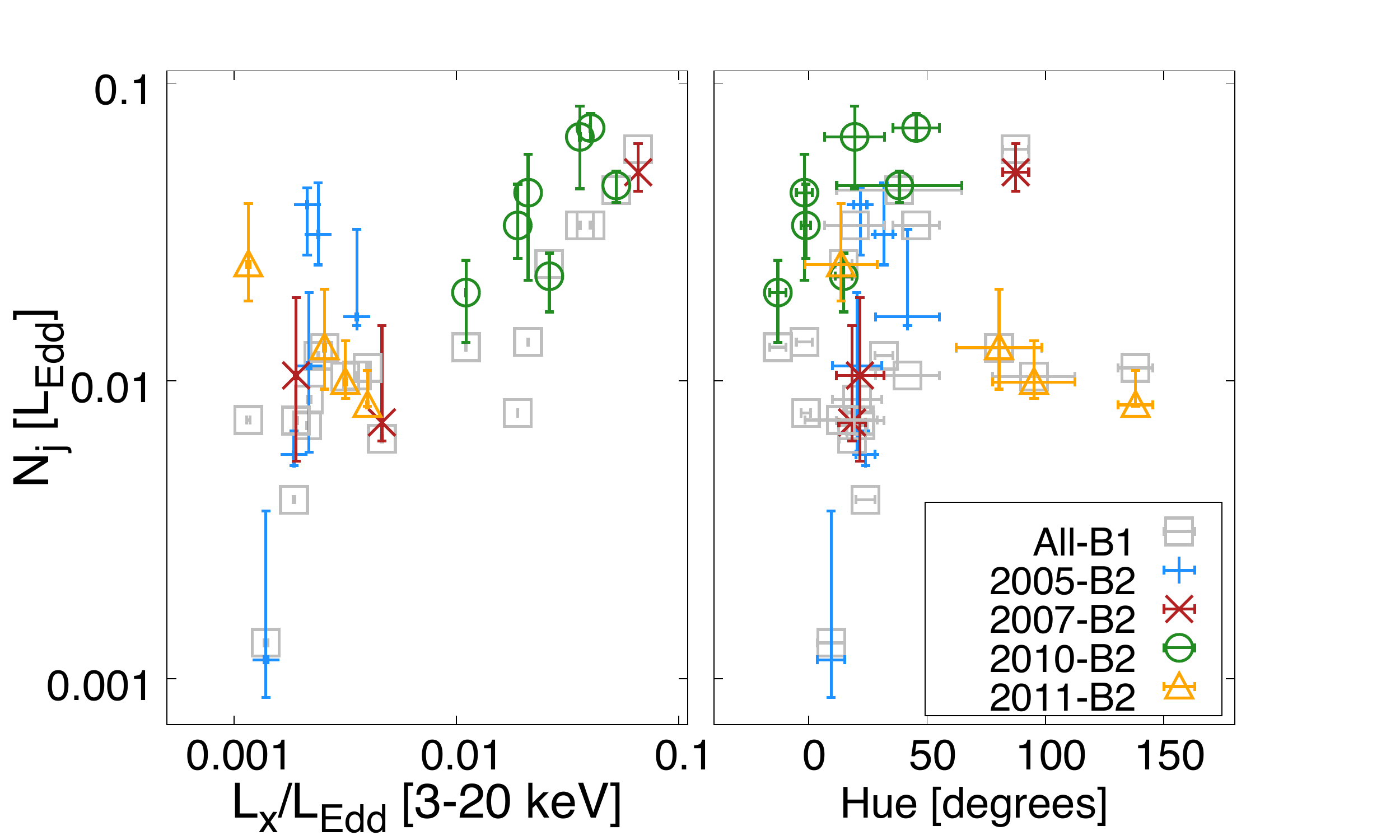}
  \includegraphics[width=0.48\linewidth]{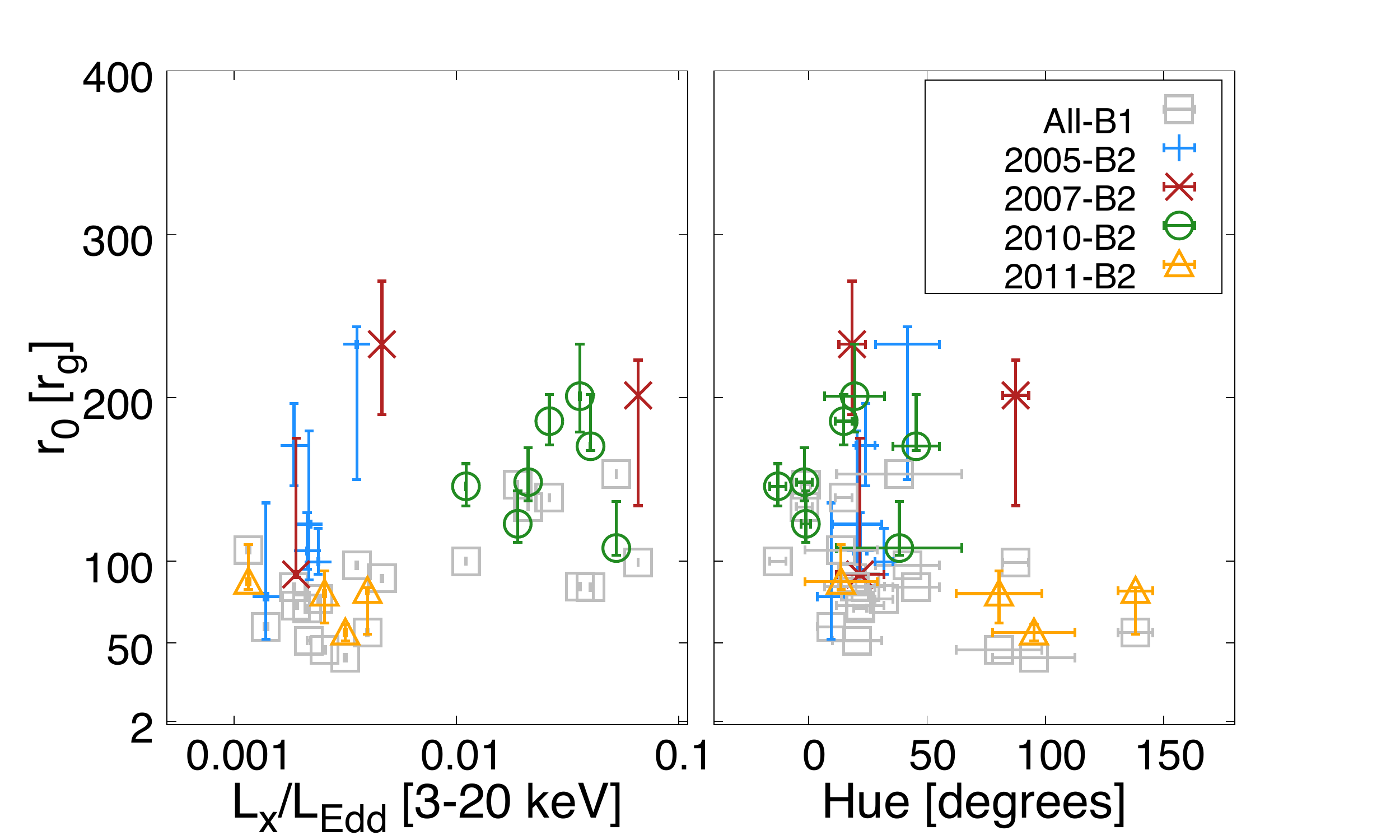}
\includegraphics[width=0.48\linewidth]{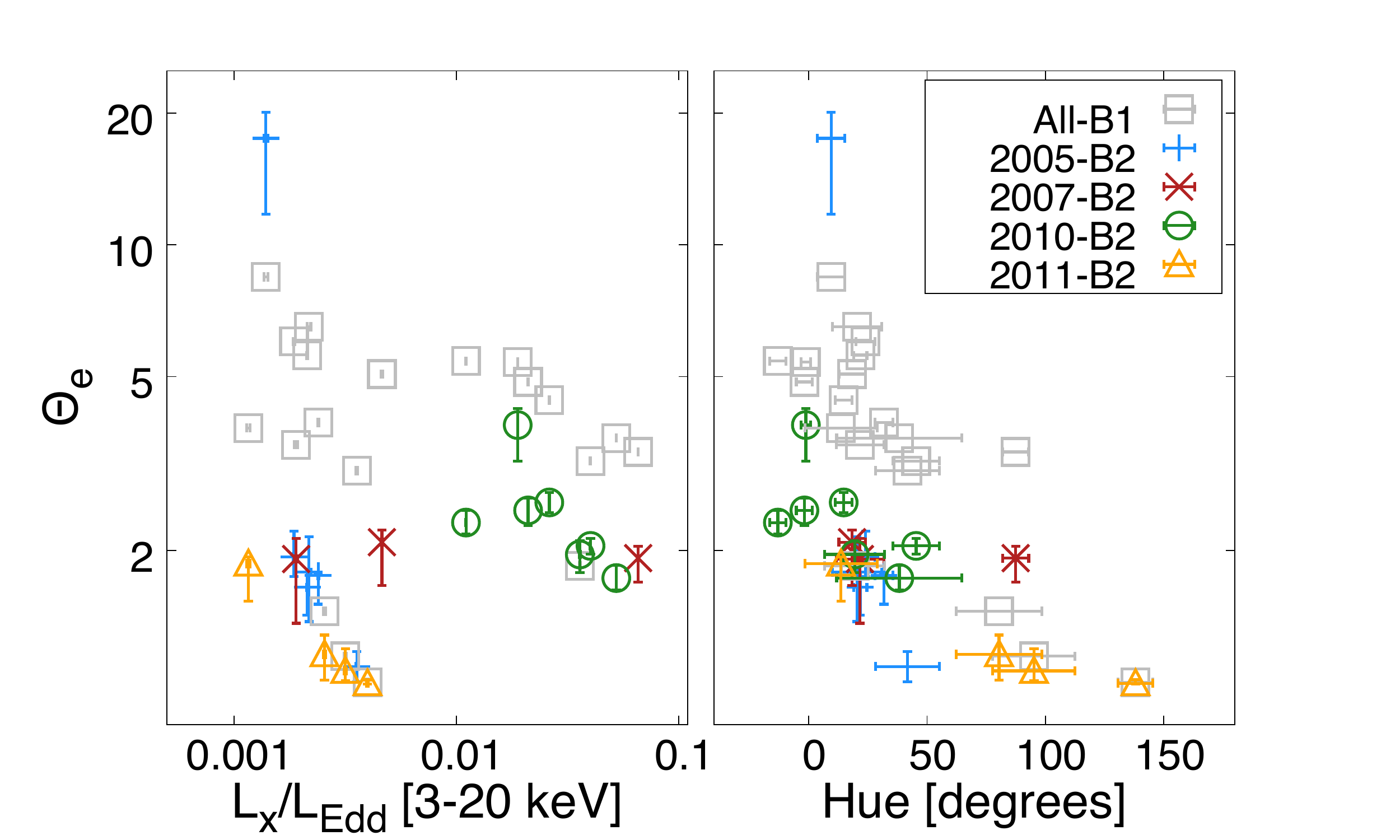}
\includegraphics[width=0.48\linewidth]{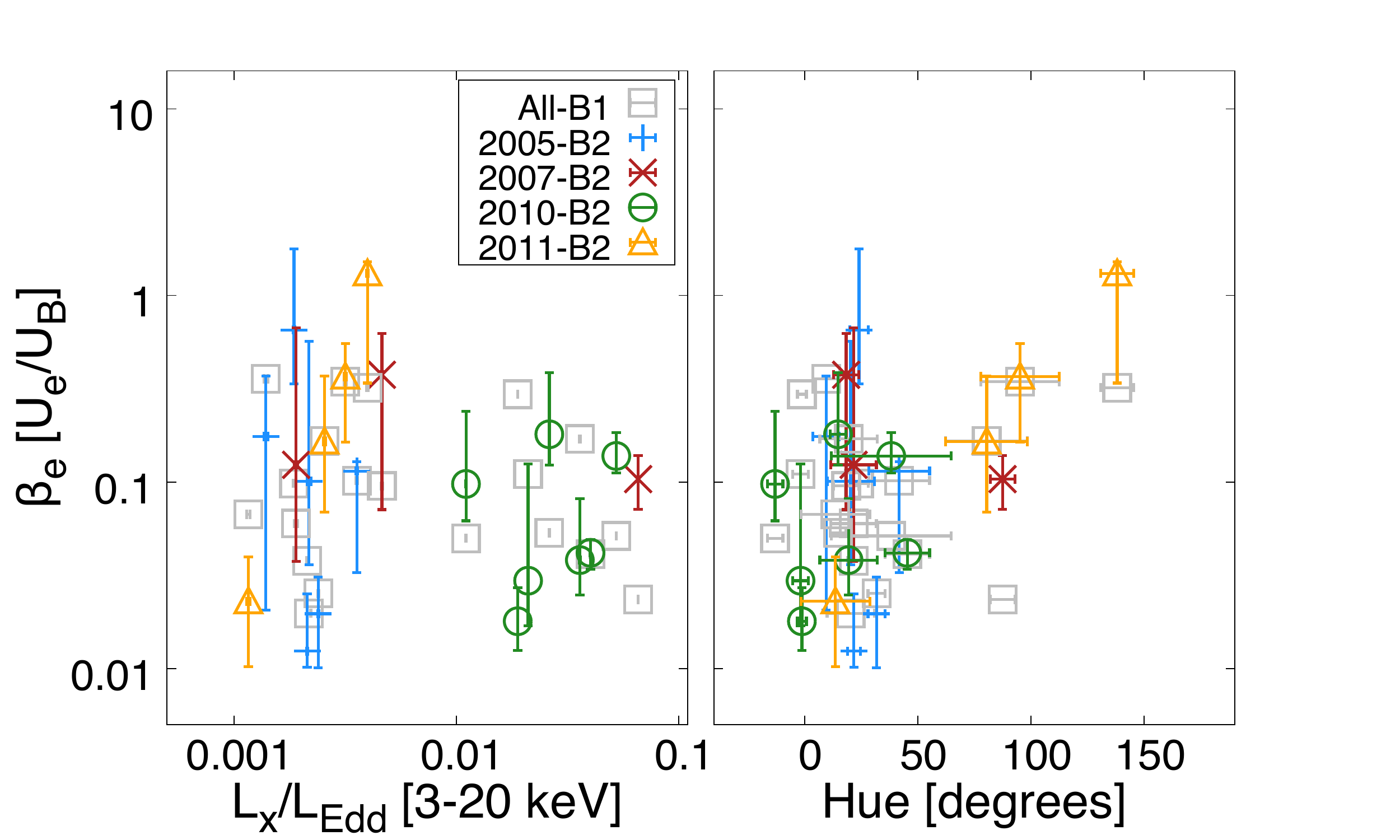}
 \caption{The MLEs (and 90\% confidence limits) of the normalised jet power, $N_{\rm j}$ (top left), jet-base radius, $r_{\rm 0}$ (top right), electron temperature, $\Theta_{\rm e}$ (bottom left), and ratio of electron to magnetic energy density, $\beta_{\rm e}$ (bottom right), as a function of Eddington-scaled X-ray data luminosity (left) and power-spectral hue (right). Gray hollow squares show the parameters derived from fits of Model~B1, and other symbols (indicated in the key) show fits of Model~B2, divided according to observation year.}
 \label{fig:jetpars}
 \end{figure*}

 Figure \ref{fig:jetpars} shows the maximum likelihood estimates (MLEs) of jet parameters $N_{\rm j}$, $r_0$, $\Theta_{\rm e}$ and  $\beta_{\rm e}$ respectively as a function of $L_{\mathrm{X}}/L_{\mathrm{Edd}}$ and power-spectral hue. Table~\ref{tab:conf} shows the numerical values of the best-fit parameters of Model~B2, and their confidence limits (we show only the best-fit values of Model~B2 as it achieves better fits to all 20 broadband spectra). 

The normalised jet power, $N_{\rm j}$, increases with increasing $L_{\mathrm{X}}$, and this is a clear trend despite the uncertainty on its value. We see that given the similar X-ray luminosities during the 2005 and 2011 outburst decays, $N_{\rm j}$ remains roughly constant as the hue decreases, until the source progresses further into the low hard state in the latter stages of outburst decay, at which point $N_{\rm j}$ decreases. This decrease is only seen in the 2005 decay, despite the similar X-ray luminosities between the 2005/2011 observations. This is likely due to the lower radio flux measured in the 2005 decay (see Figure~\ref{fig:data_fluxes}).

The jet-base radius is poorly constrained across all fits, and has a broad range from 10s to 100s of $r_{\rm g}$. There is tentative evidence for lower values of $r_0$ during the 2011 outburst decay, likely due to the degeneracy inherent between $N_{\rm j}$ and $r_0$. A decrease in $N_{\rm j}$ is constrained by decreasing radio and OIR flux, and this independent constraint on $N_{\rm j}$ is accounted for by a decrease in $r_0$ in order to fit the X-ray spectrum. 

  Despite the systematically higher values of $\Theta_{\rm e}$ when the jet IC emission dominates the X-ray spectrum, in both cases the trends are similar: $\Theta_{\rm e}$ decreases slightly with increasing power-spectral hue. This is because $\Theta_{\rm e}$ is not solely constrained by the X-ray spectrum. The hardening of the optical spectra in all 20 of our datasets is modelled by thermal synchrotron emission from the optically thin regions of the jet. The optical hardening, alongside the contribution of synchrotron emission to the radio flux at larger distances in the jet, act to constrain $\Theta_{\rm e}$. In addition, $\Theta_{\rm e}$ appears lower during the early stages of the 2011 outburst decay than in all other fits. This constraint is determined by the lower OIR fluxes (relative to radio/X-ray fluxes) in the 2011 spectra, as shown in Figure~\ref{fig:data_fluxes}. 

 We see no clear global correlation between the plasma $\beta_{\rm e}$, and the power-spectral hue, or $L_{\rm X}$, except for an apparent increase at the highest hue values, i.e., as the broadband X-ray variability is becoming narrower. This trend appears to only exist in modelling of the 2011 outburst decay, and is likely a consequence of the lower OIR fluxes relative to the X-ray flux with respect to the other multiwavelength observations (see Figure~\ref{fig:data_fluxes})---the particle energy density increases with respect to its magnetic energy density, decreasing the relative synchrotron-to-SSC contribution to the broadband spectrum. The value of $\beta_{\rm e}$ ranges between $\sim$ 0.02--1 across all the fits, with most fits yielding $\beta_{\rm e} \sim 0.1$. The trend in the fitting process is for $\beta_{\rm e}$ to be pushed to values $<1$, i.e. a magnetically-dominated jet base, which is due to increases in $N_{\rm j}$, the jet power. $N_{\rm j}$ increases in accordance with the increase in radio flux irrespective of the jet's X-ray contribution, and $\beta_{\rm e}$ in theory decreases in order to reduce the electron density in the jet base (lower electron densities lead to a lower IC flux from the jet). $\beta_{\rm e}$ is also degenerate with $r_{\rm 0}$, such that a decrease in $r_{\rm 0}$ leads to higher electron energy densities, causing $\beta_{\rm e}$ to decrease in order to redistribute the available energy density to the magnetic field, re-normalising the IC contribution to the X-rays. 
 
  In summary, we see some evidence for parameter trends that provide a physical basis for the connection between the inner accretion flow (or corona) and the jet. In particular we see distinctions between outburst rise and decay, and these changes are well tracked by the broadband X-ray variability. The jet power, $N_{\rm j}$, increases with $L_{\rm X}$ as expected. The jet base radius, $r_{\rm 0}$, is poorly constrained, with some evidence for a drop at high values of the power-spectral hue. The jet-base electron temperature, $\Theta_{\rm e}$, decreases with power-spectral hue. The ratio of electron-to-magnetic energy density shows no broad correlation, but increases with power-spectral hue and $L_{\rm X}$ in the 2011 decay.
 
 \subsection{Coronal parameter trends}
 \label{subsubsec:corona_or_jet}
\begin{figure}
 \centering
 \includegraphics[width=0.53\textwidth]{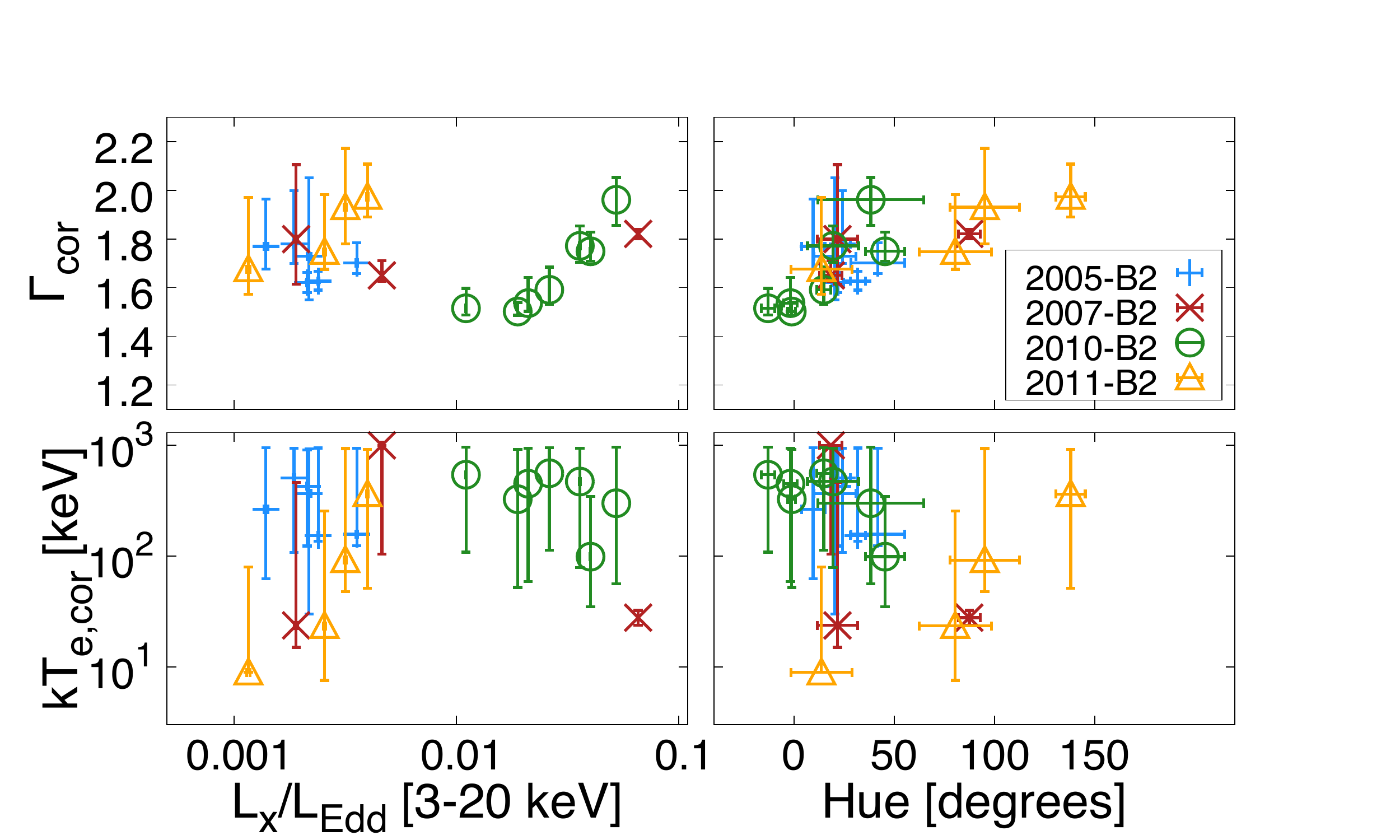}
 \caption{The MLEs (and 90\% confidence limits) of the photon index ($\Gamma_{\mathrm{cor}}$; top) and electron temperature ($kT_{\mathrm{e,cor}}$; bottom) of \texttt{nthcomp} as a function of Eddington-scaled X-ray data luminosity (left) and power-spectral hue (right). The data are divided according to observation year and thus track separate outbursts.}
 \label{fig:pars_v_hue_lx_nthcomp}
 \end{figure}

 Any trends we may expect in the coronal properties are unsurprisingly dampened by the presence of non-negligible jet contributions to the X-ray spectrum. Nonetheless some patterns exist that are worth discussing briefly. \\
 \indent Figure~\ref{fig:pars_v_hue_lx_nthcomp} shows the trends of the spectral index of the IC power-law in the corona, $\Gamma_{\mathrm{cor}}$, and the coronal electron temperature, $kT_{\mathrm{e,cor}}$, with $L_{\rm X}$ and hue. There is no observable trend between $kT_{\mathrm{e,cor}}$ and $L_{\rm X}$ or hue, any potential correlation is likely quenched by the fact that in most of the 20 GX~339$-$4 spectra the X-ray spectral coverage and photons counts are insufficient to constrain the cut-off energy, and the jet IC spectrum introduces significant scatter due to its high fractional contribution to the X-ray flux. There is a correlation between $\Gamma_{\mathrm{cor}}$ and hue and $L_{\rm X}$ during each outburst rise/decay (and striking monotonicity as a fucntion of hue, likely concurrent with X-ray hardness). Whilst a trend is expected based on our initial fits to the X-ray spectra (Section \ref{subsec:initialxrayfits}), there is added scatter in the slope again caused by the non-negligible contribution from IC emission in the jet base. 
 
\begin{table*}
 \centering
 \caption{The maximum likelihood estimates and 90\% confidence limits of fit-parameters of Model~B2 to all 20 broadband spectra of GX~339$-$4. Values are appropriately quoted to significance of the confidence limits, and thus may not match exactly the values show in Figures~\ref{fig:Rf}--\ref{fig:pars_v_hue_lx_nthcomp}. From left to right: (1) spectrum number, (2) $N_{\rm j}$, the normalised jet power, (3) $r_0$, the jet base radius, (4) $\Theta_{\rm e}$, the electron temperature in the base of the jet, (5) $\beta_{\rm e}$, the ratio of electron to magnetic energy density in the jet, (6) $T_{in}$, the inner disc temperature, (7) $R_{\rm in}$, the inner disc radius, (8) $\Gamma_{\mathrm{cor}}$, the photon index of the thermal Compton spectrum in the corona, (9) $kT_{\mathrm{e,cor}}$, the electron temperature in the corona, (10) $R_{\mathrm{f}}$, the reflection fraction, (11) chi-squared ($\chi^2$) over degrees of freedom (DoF).  }
 \begin{tabular}{@{}p{0.2cm}lllllllllr}
 \hline
 \# & $N_{\rm j}$ & $r_0$ & $\Theta_{\rm e}$ & $\beta_{\rm e}$ & $T_{\rm in}$ & $R_{\rm in}$ & $\Gamma_{\mathrm{cor}}$ & $kT_{\mathrm{e,cor}}$ & $R_{\mathrm{f}}$ & $\chi^2$/DoF\\
 & [$10^{-3}$] & [$r_{\rm g}$] & & & [keV] & [$r_{\rm g}$] & & [keV] & & \\
 \hline 
 \\
 1 & $16^{+15}_{-1}$ & $230^{+10}_{-80}$ & $1.09^{+0.09}_{-0.08}$ & $0.11^{+0.01}_{-0.08}$ & $0.30^{+0.03}_{-0.18}$ & $2^{+6}_{-0}$ & $1.70^{+0.08}_{-0.04}$ & $160^{+780}_{-30}$ & $<0.3$ & 87/68 \\ 
 
 2 & $31^{+15}_{-7}$ & $100^{+20}_{-8}$ & $1.76^{+0.03}_{-0.25}$ & $0.02^{+0.01}_{-0.01}$ & $0.19^{+0.04}_{-0.09}$ & $2.0^{+7.5}_{-0.3}$ & $1.64^{+0.04}_{-0.04}$ & $150^{+790}_{-20}$ & $<0.1$ & 106/101\\ 
 
 3 & $39^{+5}_{-13}$  & $110^{+20}_{-10}$ & $1.6^{+0.1}_{-0.2}$ & $0.013^{+0.013}_{-0.002}$ & $0.15^{+0.08}_{-0.05}$ & $5^{+5}_{-3}$ & $1.62^{+0.04}_{-0.04}$ & $400^{+500}_{-300}$ & $<0.9$ & 101/91\\ 
 
 4 & $11^{+9}_{-5}$ & $120^{+60}_{-30}$ & $1.8^{+0.4}_{-0.4}$ & $0.10^{+0.47}_{-0.07}$ & $0.18^{+0.10}_{-0.07}$ & $6^{+23}_{-5}$ & $1.7^{+0.3}_{-0.2}$ & $400^{+600}_{-300}$ & $<0.5$ & 54/30\\ 
 
 5 & $5.7^{+1.1}_{-0.5}$ & $170^{+30}_{-30}$ & $1.9^{+0.3}_{-0.2}$ & $0.7^{+1.1}_{-0.3}$ & $0.15^{+0.06}_{-0.05}$ & $3^{+3}_{-1}$ & $1.78^{+0.22}_{-0.08}$ & $500^{+400}_{-400}$ & $0.10^{+0.17}_{-0.09}$ & 63/64\\ 
 
 6 & $1.2^{+2.0}_{-0.3}$  & $80^{+60}_{-30}$ & $18^{+3}_{-6}$ & $0.13^{+0.19}_{-0.02}$ & $0.2^{+0.2}_{-0.1}$ & $10^{+0}_{-8}$ & $1.77^{+0.19}_{-0.09}$ & $300^{+700}_{-200}$ & $0.10^{+0.53}_{-0.08}$ & 16/29\\ 
 
 7 & $50^{+12}_{-7}$  & $200^{+20}_{-70}$ & $1.9^{+0.1}_{-0.2}$ & $0.10^{+0.03}_{-0.03}$ & $0.13^{+0.04}_{-0.02}$ & $6^{+4}_{-4}$ & $1.82^{+0.02}_{-0.02}$ & $30^{+5}_{-4}$ & $0.24^{+0.05}_{-0.04}$ &  351/166\\ 
 
 8 & $10^{+9}_{-5}$  & $92^{+83}_{-2}$ & $1.9^{+0.2}_{-0.5}$ & $0.12^{+0.56}_{-0.09}$ & $0.26^{+0.02}_{-0.15}$ & $3^{+7}_{-1}$ & $1.8^{+0.3}_{-0.2}$ & $24^{+440}_{-9}$ & $0.10^{+0.76}_{-0.07}$ & 38/32\\ 
 
 9 & $7^{+8}_{-1}$  & $230^{+40}_{-40}$ & $2.1^{+0.1}_{-0.4}$ & $0.4^{+0.3}_{-0.3}$ & $0.14^{+0.13}_{-0.03}$ & $10^{+0}_{-8}$ & $1.65^{+0.06}_{-0.02}$ & $>100$ & $0.10^{+0.06}_{-0.09}$ & 116/71\\ 
 
 10 & $20^{+6}_{-6}$  & $150^{+10}_{-10}$ & $2.3^{+0.2}_{-0.1}$ & $0.10^{+0.14}_{-0.04}$ & $0.23^{+0.08}_{-0.09}$ & $5^{+4}_{-3}$ & $1.52^{+0.08}_{-0.03}$ & $500^{+400}_{-400}$ & $0.05^{+0.08}_{-0.04}$ &  118/65\\ 
 
 11 & $33^{+13}_{-8}$  & $120^{+20}_{-10}$ & $3.9^{+0.4}_{-0.7}$ & $0.018^{+0.009}_{-0.005}$ & $0.3^{+0.1}_{-0.1}$ & $5^{+5}_{-3}$ & $1.50^{+0.04}_{-0.02}$ & $<900$ & $0.4^{+0.2}_{-0.2}$ & 86/67\\ 
 
 12 & $40^{+10}_{-20}$ & $150^{+20}_{-10}$ & $2.5^{+0.2}_{-0.2}$ & $0.03^{+0.10}_{-0.01}$ & $0.2^{+0.1}_{-0.1}$ & $5^{+4}_{-4}$ & $1.54^{+0.11}_{-0.03}$ & $500^{+500}_{-400}$ & $0.07^{+0.11}_{-0.06}$ & 163/65\\ 
 
 13 & $22^{+4}_{-5}$  & $190^{+20}_{-10}$ & $2.6^{+0.1}_{-0.1}$ & $0.18^{+0.21}_{-0.06}$ & $0.21^{+0.09}_{-0.09}$ & $4^{+4}_{-2}$ & $1.59^{+0.09}_{-0.06}$ & $600^{+400}_{-400}$ & $0.08^{+0.08}_{-0.06}$ & 147/67\\ 
 
 14 & $70^{+20}_{-20}$  & $200^{+30}_{-20}$ & $2.0^{+0.1}_{-0.2}$ & $0.04^{+0.04}_{-0.01}$ & $0.2^{+0.1}_{-0.1}$ & $4^{+5}_{-3}$ & $1.78^{+0.08}_{-0.07}$ & $500^{+500}_{-400}$ & $0.3^{+0.1}_{-0.1}$ & 150/65\\ 
 
 15 & $71^{+9}_{-7}$  & $170^{+32}_{-3}$ & $2.05^{+0.08}_{-0.09}$ & $0.042^{+0.007}_{-0.008}$ & $0.35^{+0.02}_{-0.14}$ & $3^{+2}_{-1}$ & $1.75^{+0.08}_{-0.04}$ & $100^{+250}_{-60}$ & $0.1^{+0.2}_{-0.0}$ & 750/66\\ 
 
 16 &  $45^{+5}_{-6}$ & $108^{+28}_{-4}$ & $1.7^{+0.1}_{-0.1}$ & $0.14^{+0.05}_{-0.03}$ & $0.17^{+0.03}_{-0.06}$ & $3^{+6}_{-1}$ & $1.96^{+0.09}_{-0.1}$ & $300^{+700}_{-200}$ & $0.03^{+0.18}_{-0.00}$ & 200/66\\ 
 
 17 & $8^{+3}_{-0}$ & $82^{+2}_{-26}$ & $1.00^{+0.02}_{-0.00}$ & $1.3^{+0.2}_{-1.0}$ & $0.18^{+0.05}_{-0.06}$ & $2.4^{+2.2}_{-0.8}$ & $1.97^{+0.13}_{-0.08}$ & $400^{+600}_{-300}$ & $0.3^{+0.3}_{-0.2}$ & 217/35\\ 
 
 18 & $10^{+4}_{-1}$  & $56^{+5}_{-7}$ & $1.06^{+0.13}_{-0.05}$ & $0.4^{+0.2}_{-0.2}$ & $0.19^{+0.09}_{-0.08}$ & $10^{+0}_{-8}$ & $1.9^{+0.2}_{-0.2}$ & $90^{+840}_{-40}$ & $0.6^{+0.4}_{-0.4}$ & 87/35\\ 
 
 19 & $13^{+7}_{-4}$  & $80^{+14}_{-18}$ & $1.2^{+0.1}_{-0.1}$ & $0.16^{+0.21}_{-0.10}$ & $0.23^{+0.06}_{-0.11}$ & $1.9^{+5.1}_{-0.2}$ & $1.75^{+0.24}_{-0.07}$ & $<200$ & $0.10^{+0.38}_{-0.09}$ & 15/34\\ 
 
 20 & $25^{+15}_{-6}$ & $88^{+23}_{-5}$ & $1.87^{+0.07}_{-0.33}$ & $0.02^{+0.02}_{-0.01}$ & $0.22^{+0.07}_{-0.11}$ & $6^{+4}_{-4}$ & $1.7^{+0.3}_{-0.1}$ & $<100$ & $0.10^{+0.54}_{-0.08}$ & 40/30\\ 
 \\
 \hline
 \end{tabular}
  \label{tab:conf}
 \end{table*}

 \subsection{Pair processes?}
 \label{subsec:pairs}
 The importance of pair processes in jet models of bright hard state BHBs, GX~339$-$4 in particular, was explored by \cite{Maitra2009}, in which a previous version to the current \texttt{agnjet} model was fit to broadband spectra of GX~339$-$4. \cite{Maitra2009} made estimates of the pair production and annihilation rates and based on those rates, adjusted their modelling to an area of parameter space in which the influence of pairs on the particle distribution and resultant spectrum were negligible. Here we expand slightly on this approach by providing a more self-consistent estimate of the energy density of pairs by numerically calculating the resultant pair distribution due to the mutual interaction of each photon field in the jet (i.e. synchrotron, SSC as well as raw and IC-scattered disc photons). 
 
\begin{figure*}
\includegraphics[width=0.49\linewidth]{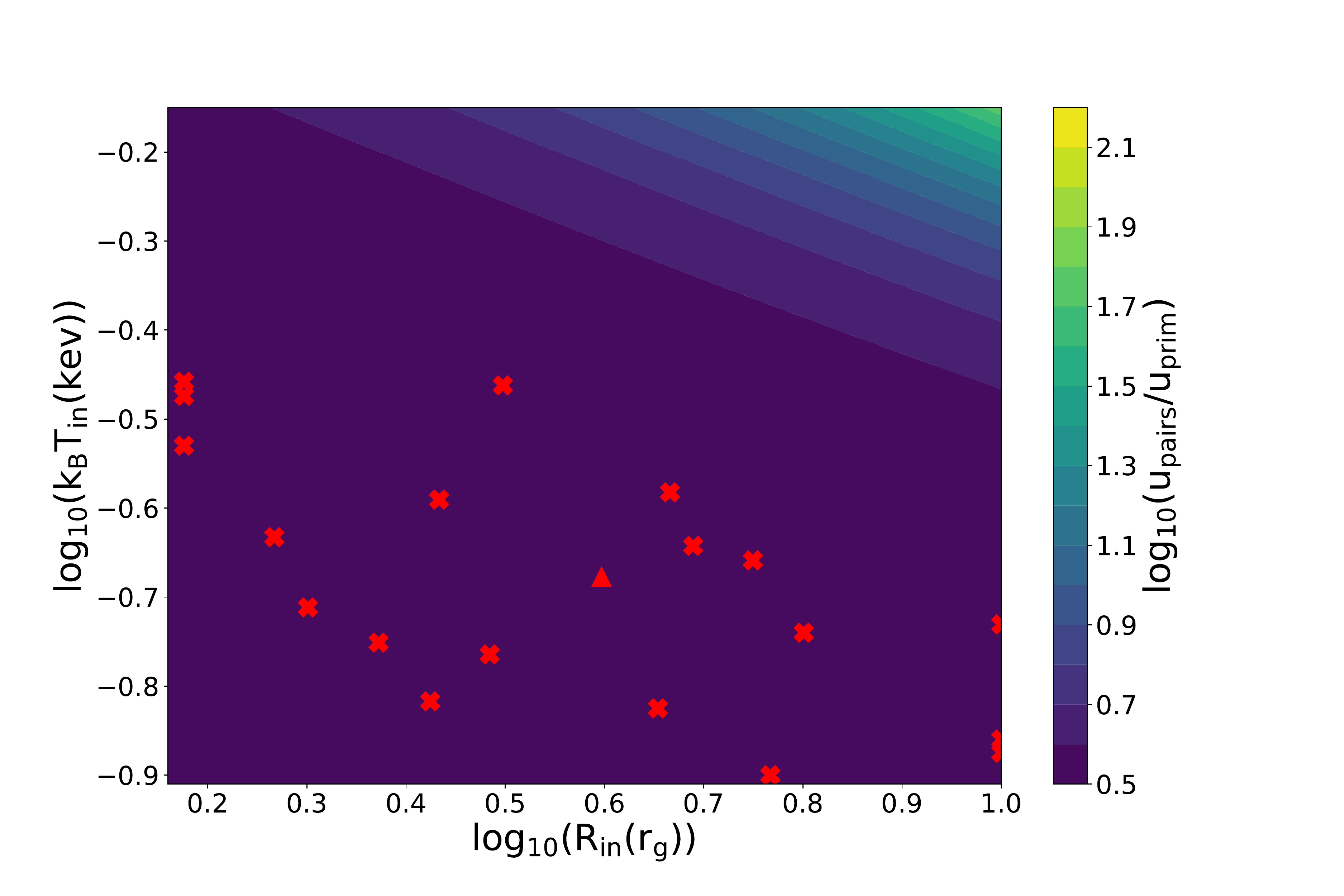}
\includegraphics[width=0.49\linewidth]{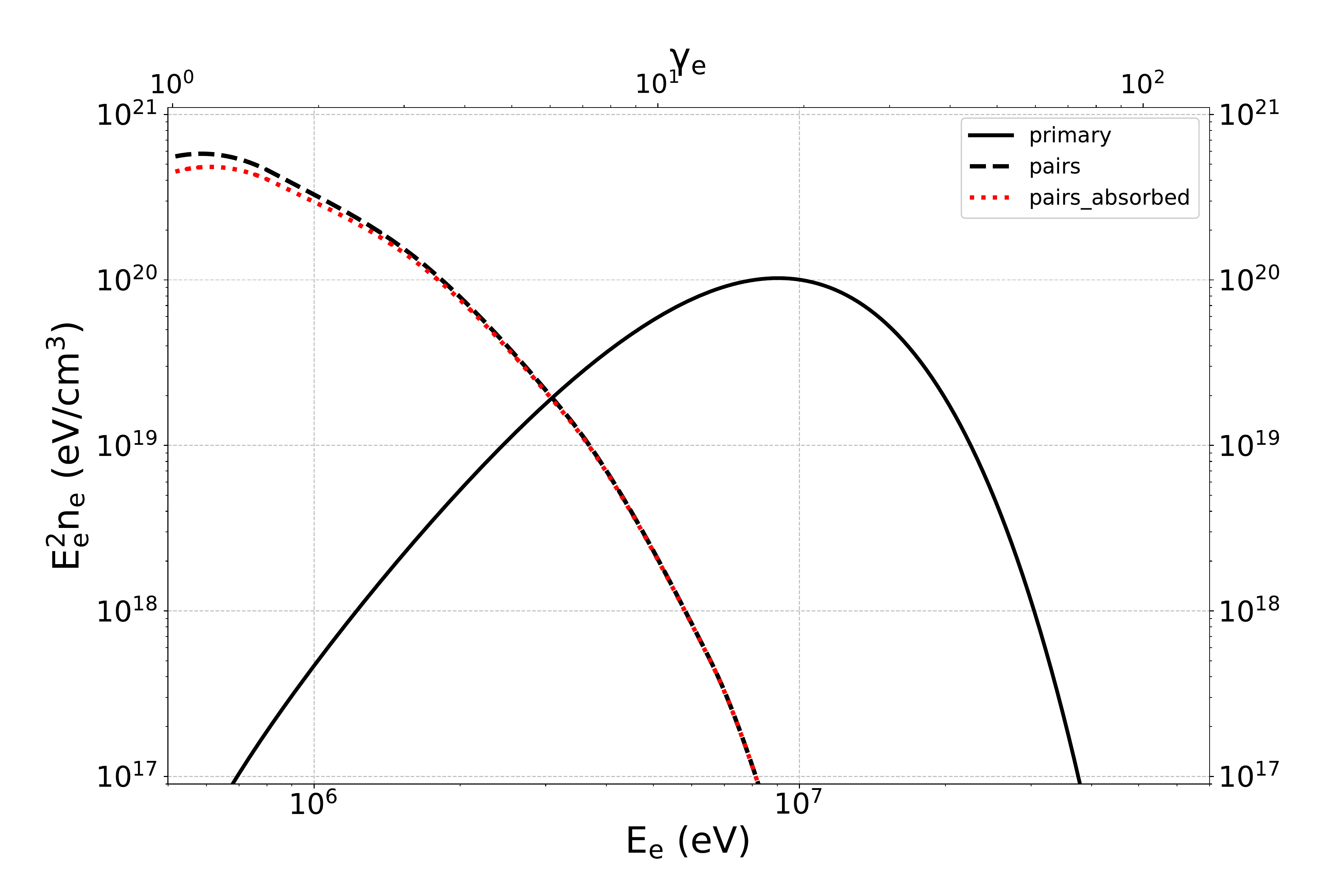}
\caption{{\bf Left:} Ratio of the energy density of electron-positron pairs to primary electrons in the base of the jet ($U_{\rm pairs}/U_{\rm prim}$), shown for a range of inner disc temperature ($T_{\rm in}$) and radius ($R_{\rm in}$). Red crosses show the values corresponding to the best fit to all 20 datasets, with a fit to a bright observation in 2010, MJD~55271 (spectrum~13), marked by the red triangle, the distribution for which is shown in the right hand figure. {\bf Right:} The energy distribution of primary electrons and secondary pairs in the base of the jet, showing the raw distribution and the absorbed one, generated using the same parameters indicated by the red triangle in the left hand figure.}
\label{fig:pairs}
\end{figure*}

 We can calculate the particle distribution self-consistently with radiative and other cooling losses balanced with the source terms, which include pair injection. Pair injection and annihilation is calculated following the formalism of \cite{Mastichiadis1995} (their Equations 57 and 60), adopting the cross-sections and production rates provided by \cite{Coppi1990}. We calculate the energy density of pairs in the base of the jet self-consistently, with synchrotron losses included. The energy density of pairs depends strongly on the most energetic photons produced by SSC in the jet, as well as IC scattering of disc photons which is seen to produce high-energy X-rays (as shown in Figure~\ref{fig:specfits}). As such, for a given set of jet parameters, the relevance of pairs may depend on the disc parameters, $R_{\mathrm in}$ and $T_{\mathrm in}$. Figure~\ref{fig:pairs} shows the ratio of the energy density of pairs in the jet base to that of the input electron energy distribution, as a function of both $R_{\mathrm in}$ and $T_{\mathrm in}$, with all 20 fit solutions marked on the plot. The energy distribution of pairs and primary electrons is then shown for one particular fit solution. One can see that for the range of best fit values found, and for the full range of $R_{\mathrm in}$ and $T_{\mathrm in}$, the energy density of pairs is comparable to the input electron energy density. However, whilst the number density is on the order of the primary number density, the average energy of the secondaries is far lower ($\gamma_e\sim1$ compared with $\gamma_e\sim20$), and closer to the non-relativistic regime, and thus they will not contribute significantly to the observed emission. As discussed clearly in Section~\ref{sec:model}, \texttt{agnjet} is dynamically dominated by its initial rest mass energy density, and so the creation of pairs in the jet, though comparable in energy density to the primary electrons, will likely not alter the dynamics significantly enough to warrant a full calculation of its effects. Such a calculation is beyond the scope of this paper, and we leave the dynamical effects of pair production in the jet to future work. 

  \section{Discussion}
 \label{sec:discussion}
 \indent Previous modelling of GX~339$-$4 with older versions of \texttt{agnjet} proposed a significant contribution in the X-ray from optically thin non-thermal synchrotron emission, either dominating the full observable X-ray band, or solely the soft band ($<10$~keV), with jet IC dominating the harder emission \citep{Markoff2003,Maitra2009}. Here we have instead considered the case in which synchrotron emission is suppressed and the jet's X-ray contribution is almost entirely dominated by thermal SSC, with some contribution from IC-scattered disc photons. There can also be contributions in the X-ray from synchrotron-emitting non-thermal electrons in the jet base or inner accretion flow, given that both are collisionless, turbulent regions in which particle acceleration can occur. \cite{Connors2017} explore this scenario in modelling of Sgr~A*, the Galactic centre supermassive black hole, and A0620-00, a BHB in quiescence, and though they both have significantly lower X-ray luminosities than GX~339$-$4 ($L_{\rm X}/L_{\rm Edd} \sim 10^{-9}$), such a scenario cannot be ruled out in the case of GX~339$-$4. However, the millisecond-to-second timescale hard X-ray lags observed in the hard state of GX~339$-$4 (see, e.g., \citealt{Nowak1999,Belloni2005,Altamirano2015}) do not favour particle acceleration as being responsible for the delayed hard X-ray emission due to the rapid timescales predicted by various particle acceleration scenarios (see, e.g, \citealt{kc15,Connors2017}). 
 
  Thus IC emission is the most likely \textit{dominant} spectral component in the X-ray. BHB hard X-ray lags in the hard state \citep{Miyamoto1988,Kazanas1997,Nowak1999} are generally interpreted as a signature of the propagation of accretion rate fluctuations in the disc responding in the coronal hard emission through IC scattering of the disc photons (e.g., \citealt{Kotov2001}). It is quite apparent that jet SSC/IC scattering off hot electrons in a low-density plasma, such as the conditions presented in the model \texttt{agnjet}, is unlikely to reproduce such lags, due to the low number of IC scatterings, and the dominance of jet synchrotron photons as the input distribution for scattering. These arguments provide both a strong qualitative and quantitative argument for IC scattering in a corona of electron temperatures in the realm of hundreds of keV with optical depths on the order of 0.1--1 as the dominant X-ray emission component in GX~339$-$4. We highlight the allusions made by \cite{Nowak2005} and \cite{Wilms2006} to the presence of multiple hard X-ray components in the low/hard state, a scenario that has already been postulated/explored for GX~339$-$4 \citep{Fuerst2015}. Recent work on spectral-timing models of BHBs in the low/hard state also postulates that there are likely two Comptonisation regions in the accretion flow \citep{Mahmoud2018}, with the only distinction from our proposed geometry being that both components are part of the gas inflow. 
 
 In Section~\ref{subsec:global_trends} we outlined that the jet is always found to be magnetically-dominated, with $\beta_{\rm e}<1$ generally holding true in all our fits. We also note that the range of the jet-base magnetisation (defined as the ratio of magnetic enthalpy to rest mass, $\sigma=B^2/4 \pi n m_p c^2$, in a force-free magnetohydrodynamic plasma such that the gas pressure is neglected) derived from the best-fit parameters always lies in the range of $\sigma=1$--$2$ (so consistently of order unity such that the magnetic field is never sub-dominant). Whilst it is important to stress that the model \texttt{agnjet} does not allow dynamically important magnetic fields (i.e. high magnetisation), we can nonetheless conclude that our modelling is dynamically consistent given the final Lorentz factors are mildly relativistic. The magnetisation necessary for  jet-launching based on recent simulations of black hole jets \citep{Tchekhovskoy2010,Tchekhovskoy2011,Sadowski2013} is typically higher (order 10). However, such simulations bias themselves toward Poynting-dominated jets due to difficulties with mass-loading into the jet. Thus given that the methods and regimes adopted by our modelling and jet-launching simulations are wholly different, it would be a misnomer to make a direct comparison.
 
  Coronal models in the context of spectral softening in BHBs predict the inward progression of the optically thick accretion disc, leading to increased cooling in the corona, and thus a lower temperature and a softer spectrum \citep{Haardt1993,Ibragimov2005}. The question of during which part of the outburst the disc has extended down to the ISCO, remains a primary discussion point for BHBs in general, none more so than GX~339$-$4. Whilst most agree that in the brightest hard states the disc extends down to the ISCO (e.g. \citealt{Gierlinski2004,Penna2010}), \cite{Miller2006} claim the disc in GX~339$-$4 sits at the ISCO throughout the low/hard state. \cite{Done2007} strongly contest this and instead claim the disc is significantly truncated and gradually moves inwards during the rise of an outburst, with an ADAF at $r < R_{\rm in}$. \cite{Kara2019} recently showed, through reverberation mapping of the X-ray emitting regions of BHB MAXI~J1820$+$070, alongside spectral modelling of the iron K line, that the coronae of BHBs are likely contracting as the source brightens in the bright hard state, whilst the disc has already reached the ISCO. As discussed in Section~\ref{subsec:global_trends}, we see evidence for decreasing jet electron temperature during the evolution of an outburst (constrained by the full broadband spectrum), but no clear trend in the corona (which is mostly constrained by the X-ray spectrum). We also find, shown in Table~\ref{tab:conf}, that $R_{\rm in}$ is likely within 10~$r_{\rm g}$ during all observations, though we stress that our constraints are weak. However, we do not find evidence for the contraction of the jet base during outburst rise or decay. We thus propose that a complete understanding of the evolution of the X-ray emitting region does need to consider the jet-corona-disc system as a whole. Though a conclusion has yet to be reached on a ubiquitous answer to this debate, it certainly appears likely that the accretion discs of BHBs are not heavily truncated during the bright hard state. 
  
  Additionally, no interpretation has yet explained why transitions between the dominant optically thin inner flow and optically thick accretion disc occur over a broad range of X-ray luminosity in BHBs \citep[$L_X/L_{Edd} \sim 0.003$--0.2]{Done2003}. Observations indicate variations in the transition luminosity within the same source, and a tendency for sources to transition at higher luminosities in outburst rise than in decay \citep{Nowak1995,Maccarone2003,Done2007} (i.e., BHB hysteresis). We have been able to track changes to the plasma conditions in the jet base whilst postulating that a separate coronal component dominates the X-ray spectrum. There are indications of a distinction between the jet properties in outburst rise and decay (at the same X-ray hardness), and these changes appear to trace the shape of the broadband X-ray variability. Thus the broadband properties of the source can point to a way to understand the hysteresis of BHBs. Since the low OIR fluxes during the onset of outburst decay (see Figure~\ref{fig:data_fluxes}) likely indicate cooler jet electrons with respect to the outburst rise, this may be further evidence for distinct plasma conditions in the inner accretion flow between the two regimes. 
  
 Multiple broadband studies of GX~339$-$4 in the hard state and across both the hard-to-soft (outburst rise) and soft-to-hard (outburst decay) state transitions have concluded that emission from the jet dominates the spectrum in the radio-to-OIR bands \citep{Corbel2002,Homan2005,Russell2006,Gandhi2008,Coriat2009,Gandhi2011,Buxton2012}, and even perhaps into the UV bands \citep{Yan2012}. However, the nature of the emission is still uncertain. Whilst some claim the jet synchrotron break occurs in the mid-Infrared ($>10^{13}$~Hz; \citealt{Gandhi2011}), others conclude that the optically thick portion of the jet spectrum extends from the radio to beyond the V band ($>10^{14}$~Hz; \citealt{Coriat2009,Dincer2012,Buxton2012})---we note here that these conclusions are not all based on the same observations, and we may expect differences in the break location during different outbursts. A bias exists in our modelling, since we have fixed the location of particle acceleration in the jet, $z_{\rm acc}$. However, our results show that the flatter portion of the lower-frequency IR spectra and the bluer portion of the optical spectra can be modelled as a superposition of thermal and non-thermal jet synchrotron components, where the break frequency is always situated below the OIR bands. It should be noted that although several of our fits fail to capture the indices of the OIR spectra, in many cases the combination of thermal and non-thermal synchrotron emission can easily conspire to hide the jet break in the observed spectrum and successfully reproduce the optical up-turn. 
 
 We also find that during the decay of the 2011 outburst the jet break should be more pronounced due to the OIR dip relative to the radio flux, and the inversion of the OIR spectrum is well-modelled by thermal synchrotron. We cannot rule out the contribution from disc reprocessed emission during the soft-to-hard transition, but at these low X-ray luminosities ($\le 0.01~L_{\mathrm{Edd}}$) the jet spectrum is most likely to be dominating in the optical \citep{Gandhi2008}. Evidence for the optical emission of BHBs being dominated by synchrotron radiation in the jet within $\sim10^3~r_\mathrm{g}$ of the accreting black hole has now been seen in several BHBs, XTE~J1118$+$480 \citep{Kanbach2001}, GX~339$-$4 \citep{Gandhi2008,Gandhi2011}, the recently discovered transient MAXI~J1820$+$070 (see, e.g., \citealt{Townsend2018}) and V404~Cygni \citep{Gandhi2017}, with V404~Cygni showing confirmed activation of the self-absorbed radio jet alongside the onset of rapid optical variability. Our comprehensive modelling of GX~339$-$4 during both the rise and decay of multiple outbursts provides supporting evidence for a physical picture in which the jets of BHBs dominate the broadband spectrum at radio-to-OIR frequencies, and thus likely also contribute a non-negligible X-ray flux.   
 
  In a simplistic framework in which the corona is an outflowing, purely non-thermal plasma, to successfully explain the trend of increasing reflection fraction ($R_{\mathrm{f}}$) with X-ray power-law spectral slope ($\Gamma_{pl}$), we expect the bulk velocity of the corona ($\beta_j$) to decrease with increasing luminosity (see, e.g., \citealt{Beloborodov1999}). As noted by \cite{Done2007}, this disagrees with fundamental observations of BHB jet radio cores \citep{Fender2006}, where higher bulk velocities are observed at higher luminosities. However, a more complete outflow model with a physical connection between the bulk flow properties and dissipation of energy into the radiating electrons (beyond the physical jet model put forward in this work) points to other scenarios in which the correlation between $R_{\mathrm{f}}$ and $\Gamma_{pl}$ can be realised without violating requirements on the jet dynamics. 
  
 For example, in \texttt{agnjet} the electrons energies are in a Maxwell-J\"{u}ttner distribution with initial temperatures $\Theta_{\rm e} \ge 1$, and remain quasi-isothermal, cooling only in proportion to the jet acceleration in the z-direction ($T(z) = T_0[\gamma_j(z) \beta_j(z)]^{1-\Gamma}$, where $\Gamma = 4/3$ is the adiabatic index. The electrons in the outer regions of the jet must remain hot ($\Theta_{\rm e} \ge 1$) to reproduce the flat/inverted radio spectral index (and in our modelling particle acceleration occurs, so further energy has been dissipated into the electrons), but the electrons in the jet base may have low initial temperatures typical of coronae ($\Theta_{\rm e} \sim 0.2$), and heating can occur rapidly due to turbulence, shocks, thermal conduction or magnetic reconnection \citep{Quataert2000,Johnson2007,Sironi2015,Ressler2015,Rowan2017}. Our work here shows the importance of such a model. For example, the apparent decrease in jet-base electron temperature ($\Theta_{\rm e}$) with increasing power-spectral hue, i.e., as the source progresses through the hard state, agrees with the general consensus that as BHBs evolve through their outbursts the corona is cooling and becoming more compact \citep{Haardt1993,Ibragimov2005}. As discussed already, an unanswered question still exists as to the evolution of the coronal-disc setup, despite a recent breakthrough indicating that the disc remains close to the ISCO in the bright hard state \citep{Kara2019}. We argue that developing a clearer idea of how the corona and the jet interact may be a critical stepping stone in understanding the co-evolution of both with the accretion disc.
 
 \section{Summary and conclusions}
 \label{sec:conclusions}
We have combined a thermal IC-scattering corona (\texttt{nthcomp}: \citealt{Zdziarski1996,Zycki1999}) and a jet in broadband spectral modelling of GX~339$-$4, with two fundamental differences between the two IC scattering treatments: the input soft-photon distribution for the jet IC scattering (SSC + IC scattering of disc photons) in \texttt{agnjet} is dominated by thermal synchrotron photons, and the electrons are strictly relativistic ($\Theta_{\rm e} \ge 1$, $kT_e \ge 511~\mathrm{keV}$) within a plasma of low optical depth ($\tau\sim10^{-4}$--$10^{-2}$), whereas the input photons of the corona in \texttt{nthcomp} are disc blackbody photons at $T_{\mathrm{BB}} \sim$ 0.01--1 keV, and the electrons are typically on the order of $kT_e \sim10$s--$100$s of keV in a plasma of higher optical depth ($\tau\sim0.1$--$1$). Analogies to such a physical model can be found in many simulations of black hole accretion flows in which the inner flow is ADAF-like (geometrically-thick and optically thin) and the jet is a Poynting-dominated (we find jet-base magnetizations of order unity), low-density funnel launched via the Blandford-Znajek mechanism (see, e.g., \citealt{Mckinney2006,hk06,Tchekhovskoy2010,Tchekhovskoy2011,Sadowski2013}, and references therein). Only the cooler, higher optical depth coronal component of \texttt{nthcomp} can successfully reproduce the X-ray spectra of all 20 GX~339$-$4 datasets we modelled, and this is due to precisely the two identified model discriminants described. Given these conditions, the main results of our comprehensive modelling of GX~339$-$4 can be summarised in the following points: 
 
 \begin{itemize}
 
 \item Even if IC scattering in the corona dominates the X-ray spectrum of GX~339$-$4 in the low/hard state, there will still likely be a non-negligible contribution from jet IC-scattered photons. 
 
\item There are trends in the physical properties of the jet during both outburst rise and decay, even with the presence of a dominant coronal component, and these changes appear to show correlations with the shape of the broadband X-ray variability. 

\end{itemize}

Addressing the former conclusion first,  we find ratios of jet-to-corona continuum flux of a few to $\sim50\%$ in the 3--100~keV band across all fits. However we note that this conclusion is strongly model-dependent. The jet (\texttt{agnjet}) electrons are treated relativistically in a plasma at low optical depths ($\tau \le 0.01$). A treatment which includes cooler electrons in a region of higher optical depth, producing IC spectra with less curvature, would likely reduce the difference in spectral shape between the corona and jet base IC emission in our modelling (and in fact may return to a scenario where the `corona' is synonymous with the base of the jet). We cannot rule out a contribution to the X-ray from non-thermal optically thin jet synchrotron emission \citep{Markoff2003,Maitra2009}. We have artificially suppressed such a contribution in order to limit the degeneracies in our modelling, along with a strong argument for its non-dominance as a contribution in the X-ray emission of GX~339$-$4 (and other BHBs in the hard state)---a mix of synchrotron and IC jet emission, dominating the soft and hard X-rays respectively, struggles to explain the ubiquitous presence of hard X-ray lags (see, e.g., \citealt{Nowak1999,Belloni2005,Altamirano2015}). \\
\indent On the latter conclusion, by tracking the jet and coronal parameters as a function of both X-ray luminosity and the power-spectral hue (a simple characteriser of the shape of the broadband X-ray rms variability), we have shown some trends appear in the jet properties. As is expected, the jet power increases with X-ray luminosity, constrained primarily by the observed quasi-simultaneous radio flux. The jet-base electron temperature, $\Theta_{\rm e}$, can be seen to slightly decrease with increasing hue, thus coincident with the strengthening and narrowing of the broadband X-ray variability. The jet base is more compact with cooler electrons during the 2011 outburst decay of GX~339$-$4 as the shape of the X-ray variability strengthens and narrows. At lower values of the power-spectral hue, when the X-ray variability has a broader shape, we see no clear distinctions in the jet physics between outburst rise and decay. Our results point to a way of constraining the geometrical changes by linking the evolving X-ray variability in the inner regions to the plasma conditions further out in the jet.\\
 \indent Determining the contribution of jet emission in the X-ray still remains a difficult task in the modelling of BHBs. The jet contribution must be quantified in order to better constrain the fraction of hard X-ray emission reflected off BHB accretion discs \citep{Ross1999,Ross2005,Dauser2010,Garcia2014,Garcia2015_2}, since if a significant fraction of the X-rays are beamed away from the disc, the emissivity profile along the disc is affected, and therefore the reflection fraction changes \citep{Dauser2013,WilkinsGallo2015} and the features relevant for determining the black hole spin and inner disc radius are altered. We will address the importance of the jet contribution to X-ray disc reflection in a forthcoming paper (Connors et al., in preparation). \\
 \indent A significant caveat that all jet models so far suffer from is that the plasma conditions which determine the spectrum of the jet are disconnected from the jet dynamics. With \texttt{agnjet} for example, the velocity, particle density, and magnetic field profiles are pre-calculated dynamical quantities in the model, and the broadband spectrum follows from the radiative calculations, with cooling effects only incorporated into that resulting spectrum. An improved treatment would involve reducing the number of free parameters by physically linking the radiative calculations with the jet dynamics. Self-similar MHD solutions of a relativistic jet presented by \cite{Ceccobello2018}  (building on work by \citealt{Polko2010,Polko2013,Polko2014}) provide the groundwork for such a treatment. By combining the plethora of dynamical jet solutions presented by \cite{Ceccobello2018} with radiative calculations such as those presented in this work, we shall in future be able to find more physically-realistic solutions for a given system (BHB or AGN) and perform model-fitting to retrieve more meaningful results with less degeneracies.

 \section*{Acknowledgements}
 We thank the referee for their useful comments which have gone a long way to improving this manuscript. DK thanks Maria Petropoulou for informative discussions regarding pair processes. RMTC also thanks Javier Garcia for useful discussions. 
 
This research has made use of \texttt{ISIS} functions provided by ECAP/Remeis observatory and MIT (http://www.sternwarte.uni-erlangen.de/\texttt{ISIS}/). 

RMTC is thankful for support from NOVA (Dutch Research School for Astronomy), and acknowledges funding from NASA grant No. No. 80NSSC177K0515. CC acknowledges support from the Netherlands Organisation for Scientific Research (NWO), grant Nr. 614.001.209. SM, DK, and ML acknowledge support from NWO VICI grant Nr. 639.043.513. VG is supported through the Margarete von Wrangell fellowship by the ESF and the Ministry of Science, Research and the Arts Baden-W\"urttemberg.

\bibliographystyle{mnras}
\bibliography{references}

\label{lastpage}
\end{document}